\documentclass[usenatbib]{mnras}

\usepackage{hyperref}

\usepackage[T1]{fontenc}
\usepackage{ae,aecompl}

\usepackage{graphicx}	
\usepackage{amsmath}	
\usepackage{amssymb}	

	\newcommand{\del}[1]{\partial_#1}
	\newcommand{\Dr}[1]{\frac{d}{d#1}}
	\newcommand{\DDr}[1]{\frac{d^2}{d{#1}^2}}
	\newcommand{\Req}[0]{R_\mathrm{e}}
	\newcommand{\Rc}[0]{R_\mathrm{c}}
	\newcommand{\ROL}[0]{R_\mathrm{OL}}
	\newcommand{\Rmatch}{R_\mathrm{m}}
	\newcommand{\Rw}{R_\mathrm{w}}
	\newcommand{\epsC}[0]{\varepsilon_\mathrm{c}}
	\newcommand{\epsOL}[0]{\varepsilon_\mathrm{OL}}
	\newcommand{\eNap}[0]{\mathrm{e}}
	\newcommand{\sigr}[0]{\sigma_\mathrm{r}}
	\newcommand{\sigi}[0]{\sigma_\mathrm{i}}
	\newcommand{\jc}[0]{j_\mathrm{c}}
	\newcommand{\OmegaO}[0]{\Omega_\mathrm{o}}

\title[Role of corotation radius in the low $T/W$ dynamical instability]
{On a role of corotation radius in the low $T/W$ dynamical instability of differentially rotating stars}
\author[S. Yoshida and M. Saijo]
{Shin'ichirou Yoshida$^{1}$
\thanks{E-mail: yoshida@ea.c.u-tokyo.ac.jp}
and Motoyuki Saijo$^{2}$
\thanks{E-mail: saijo@aoni.waseda.jp}
\\
$^{1}$ Department of Earth Science and Astronomy, Graduate School of Arts and Sciences, The University of Tokyo, \\Komaba 3-8-1, Meguro-ku, Tokyo 153-8902, Japan\\
$^{2}$ Department of Physics, Waseda University, Shinjuku, Tokyo 169-8555, Japan}
\date{Accepted 2016 November 22. Received 2016 November 22; in original form 2016 October 13}
\volume{}
\pubyear{201x}
\begin{document}
\label{firstpage}
\pagerange{\pageref{firstpage}--\pageref{lastpage}}
\maketitle
%
\numberwithin{equation}{section}
%
\begin{abstract}
We investigate the nature of so-called low $T/W$ dynamical instability in a differentially rotating star by focusing on the role played by the corotation radius of the unstable oscillation modes. An one dimensional model of linear perturbation, which neglects dependence of variables on the coordinate along the rotational axis of the star,
is solved to obtain stable and unstable eigenmodes.
A linear eigenmode having a corotation radius, at which azimuthal pattern speed of the mode coincides with the stellar angular velocity, is categorized to either a complex (growing or damping) mode
or a purely real mode belonging to a continuous spectrum of frequency.
We compute canonical angular momentum and its flux to study eigenmodes with corotation radius. 
In a dynamically unstable mode, sound wave transports its angular momentum in such a way that the absolute value of the angular
momentum is increased on both sides of the corotation radius. 
We further evaluate growth of amplitude of reflected sound wave incident to a corotation
point and find that the over-reflection of the wave and the trapping of it between the corotation radius and the
surface of the star may qualitatively explain dependences of eigenfrequencies on the stellar differential rotation. The results suggest that the low $T/W$ instability may be caused by over-reflection of
sound waves trapped mainly between the surface of the star and a corotation radius.

\end{abstract}


\section{Introduction}
Stars are naturally rotating and potentially exhibit non-axisymmetric instabilities.
Within linear perturbation analysis non-axisymmetric dynamical instability of rotating stars have been
mainly explored for uniformly rotating cases (see, however, \citet{Toman_etal1998, Karino_Eriguchi2003}). 
Here we classify dynamical instability of a system as an instability in which the system becomes unstable without any dissipative processes, in contrast to secular instability in which a dissipation process in the system is essential. A bar-mode oscillation ($m=2 ~f$-mode, where $m$ is the azimuthal wave number of scalar axial harmonics $\eNap^{im\varphi}$) is the most susceptible to destabilization by stellar rotation and studied in detail.  The mode in Newtonian stars is shown to have a dimensionless quantity $T/W\sim 0.27$ for neutral stability \citep{Tassoul_Ostriker1968, Ostriker_Tassoul1969, Ostriker_Bodenheimer1973, Tassoul1978book}, where $T$ and $W$ are rotational kinetic energy and gravitational binding energy of a star respectively. This parameter $T/W$
measures how rapid an object is spinning. 
Relativistic gravity reduces the critical value of $T/W$ for the instability to $\sim 0.24-0.26$, depending on the compactness of the star \citep{Shibata_etal2000barmode, Saijo_etal2001, Baiotti_etal2007}.
A mechanism of the classical bar-mode instability is explained as merger of two $f$-modes to results in a pair of complex conjugate modes \citep{Schutz1980_pap3}.

For differentially rotating stars another type of dynamical instability has been found through numerical hydrodynamic simulations \citep{Picket_etal1996, Centrella_etal2001, Shibata_etal2002, Shibata_etal2003, Saijo_etal2003, Ou_Tohline2006, Corvino_etal2010}. 
The new instability is called `low $T/W$' dynamical instability since the critical value of $T/W$ is 
significantly smaller than that of classical bar-mode and may be as low as ${\cal O}[10^{-2}]$. 
The instability is also seen in various values of $m$ and of stiffness of stellar matter,
but the saturation amplitude of this instability seems to be smaller than that of the classical bar-mode.
This characteristics of the instability is advantageous for detection of gravitational wave from it, once a compact star is susceptible to the instability. 
\citet{Shibata_etal2002} showed in bar unstable mode ($m=2$) that the gravitational waveforms persists for 
hundreds of central rotation periods
without being disrupted by non-linear hydrodynamic effects.

On the other hand one-armed spiral mode ($m=1$) in low $T/W$ instability may have a wave-packet like waveform in contrast to the unstable bar mode 
\citep{Saijo_etal2003}, and it is also discussed in the context of post-bounced core of a core-collapse type supernova \citep{Ott_etal2005, Kuroda_etal2014} and of binary neutron star mergers \citep{Lehner_etal2016, Radice_etal2016}
(see also \citet{Takiwaki_etal2016} for the role of one-armed spiral instability in neutrino radiation of
core-collapse supernova explosion). 

The mechanism of the low $T/W$ dynamical instability, on the other hand, has not been fully investigated.
\citet{Watts_etal2005} argued that the low $T/W$ dynamical instability is a shear instability which appears when the degree of differential rotation of a star
is sufficiently high and $f$-mode develops a corotation radius at which the azimuthal pattern speed of the
mode coincides with the angular frequency of the star.
\citet{Saijo_Yoshida2006} succeeded in employing canonical angular momentum of linear perturbation
to classify the different unstable modes by analysing its behaviour around the corotation radius. They combined a hydrodynamical simulation and a linear perturbation analysis in frequency domain to elucidate differences among low $T/W$ dynamical instabilities of $m=1$, $m=2$ and classical bar mode instability.
\citet{Passamonti_Andersson2015}, by a linear perturbation analysis in time domain, confirmed that 
the $\ell=m=2$ $f$-mode ($\ell$ is the degree of associated Legendre function $P_\ell^m$)
is destabilized when it develops a corotation point. They also showed the imaginary part of eigenfrequency becomes larger as the mode is deeper inside the corotation band of frequency and proposed an empirical formula for the imaginary part of the unstable $f$-mode eigenfrequency as a function of $T/W$ value and the degree of differential rotation. 

Although we learned from these works an indication that corotation radius inside the star seems to play an essential role to this instability, still detailed exploration is required to understand a physical mechanism that causes the instability.
What is the role played by the corotation radius in the low $T/W$ dynamical instability, if it actually induces the instability?
Does an eigenmode having a corotation radius always become unstable? 
What are differences/similarities among the low $T/W$ dynamical instability and other known shear instabilities in
stars and discs?
All these issues of principles are yet to be answered. In view of the importance of the instability in astrophysics of compact stars, these issues need to be tackled.
For accretion discs and tori in which differential rotation is inevitable eigenmode and stability analyses have been developed quite well (for a review of thin disc oscillations, see \citet{Kato2001}). Thick discs or
tori are also susceptible to dynamical instabilities \citep{Papaloizou_Pringle1984, Balbinski1984a, Balbinski1985, Drury1985, Goldreich_Narayan1985, Blaes_Glatzel1986, Goldreich_etal1986_papI, Goodman_etal1987_papII, Narayan_etal1987, Luyten1990}. In fact we here study low $T/W$ dynamical instability being inspired by these precedent studies of instabilities in discs and tori, especially by \citet{Drury1985} and \citet{Tsang_Lai2008}. They studied how a wave travelling in a accretion disc or torus incident to a corotation radius is over-reflected so that the reflected wave have a larger amplitude than the incident one. 

In this paper we study by linear perturbation characteristics of the low $T/W$ dynamical instability of differentially rotating stars in Newtonian gravity with corotation, focusing on equatorial motions of the fluid.
First we detail normal modes of differentially rotating stars.  \citet{Saijo_Yoshida2016} studied the unstable normal modes in the same situation to ours, finding that the existing pulsation modes become unstable due to the existence of corotation radius inside the star.  The feature of the unstable mode eigenfrequency and its eigenfunction in the linear analysis roughly agrees with that in three dimensional hydrodynamical simulations in Newtonian gravity.  Although they only focus on the unstable normal modes, we extend their study to the normal modes including those in continuous spectrum of eigenfrequency \citep{Yoshida2006}, which cannot be tackled by their analysis, and stable damping modes to understand the whole physical picture around corotation.
Second we study angular momentum transport around the corotation radius, which leads to the instability.  \citet{Saijo_Yoshida2006} find that the canonical angular momentum seems to flow inward through corotation in the linear unstable regime of evolution from the analysis of canonical angular momentum density.  
We focus on the canonical angular momentum flux to understand such effect in terms of the role of corotation radius.
Combining above two results in terms of corotation, we finally propose and demonstrate one physical mechanism to generate this instability; an over-reflection of sound wave at the corotation radius.  We examine this proposal for unstable eigenmodes.
 
This paper is organized as follows. In section \ref{sec:basic equations} our formulation of linear analysis 
in the equatorial motion is introduced.
In section \ref{sec: Results} we present the results of our analysis. First the structure of eigenmode spectrum with a possible presence of a corotation radius is detailed. Second we focus on the modes  having corotation radii and characterize them with canonical angular momentum density and its flux. We also look at energy flux of sound wave at the corotation radius. Finally we argue that low $T/W$ dynamical instability with $m\ge 2$ may be realized by the over-reflection of sound wave 
at the corotation radius, which is trapped between the surface and the corotation radius.
In section \ref{sec: Discussions} we discuss a possible origin of the instability 
and conclude our paper in section \ref{sec: Conclusions}.
In the appendices, we briefly present two potential formalism of linear perturbation, a series solution around the corotation radius and the definition of canonical angular momentum density and its flux.

In this paper we use the Latin indices $a,b,\dots$ for components of spatial vectors and tensors. The Einstein's rule of  contraction of indices is adopted.

\section{Formulation}
\label{sec:basic equations}
\subsection{Basic equations}
The equations we solve are the continuity of mass equation, the Euler equation of fluid and the Poisson equation for
gravitational potential. The continuity equation is written as
	\begin{equation}
		\del{t}\rho + \nabla_a(\rho v^a) = 0.
		\label{eq:continuity0}
		\end{equation}
Here $\rho$ and $v^a$ are mass density and velocity of fluid. $\nabla_a$ is the covariant derivative compatible with a
spatial metric $\gamma_{ab}$.
We assume that the stellar fluid is barotropic and its equation of state (EOS) is polytropic. 
Thus we have a neutrally stable configuration to convectional instability. 
In this case, the Euler equation is written as
	\begin{equation}
		\del{t} v_a + v^b\nabla_bv_a = -\nabla_a U,
		\label{eq:Euler0}
		\end{equation}
where a scalar potential $U$ is defined as
	\begin{equation}
		U = h + \Phi,
		\end{equation}
by the enthalpy $h=\int dp/\rho$ and the gravitational potential $\Phi$. 
%
The gravitational potential is
a solution of the Poisson equation,
	\begin{equation}
		\nabla^a\nabla_a\Phi = 4\pi G\rho.
		\label{eq:Poisson0}
		\end{equation}

\subsection{Linearised equations}
Equations (\ref{eq:continuity0}), (\ref{eq:Euler0}) and (\ref{eq:Poisson0}) are linearised around an equilibrium state. Since an equilibrium state
we are interested in is stationary and axisymmetric, the perturbed quantities as $\delta U$ are assumed to be expressed by a single harmonic component in time $t$ and azimuthal coordinate $\varphi$ as $\delta U\propto \eNap^{-i\sigma t + im\varphi}$. 
Hereafter angular velocity and frequency are normalized by $\OmegaO=\Omega(R=0)$ as 
$\Omega/\OmegaO$ and $\sigma/\OmegaO$.
To write down our basic system of linearised equations, we adopt two potential formalism by \citet{IpserLindblom2pot1990}. Linear adiabatic perturbations to a self-gravitating stationary 
and axisymmetric star (disc) may be described by two scalar potentials $\delta U=\delta h + \delta\Phi$
and $\delta\Phi$, which solve a couple of second order elliptic differential equations of coordinates in the meridional plane.
Here $\delta h$ is perturbation to enthalpy, which is defined as $\delta h=\delta p/\rho$, and $\delta\Phi$ is gravitational potential. As for the meridional coordinate, we adopt the cylindrical coordinate $(R,z)$.  A derivation of the equations for $\delta U$ and $\delta\Phi$ are given in appendix~\ref{appendix:2potentials}.

We introduce a model to our problem which makes the equations simpler and easier for physical
characteristics of eigenmodes to be extracted. The model we call cylindrical model
\citep{Saijo_Yoshida2016}
deals with a system of equation only in the equatorial plane and discards the $z$-component of vectors
as well as $z$-dependence of the perturbed variables. We naturally choose cylindrical coordinate $(R,\varphi,z)$
to write down basic equations. \citet{Saijo_Yoshida2016} utilizes the model and obtain results of a qualitative agreement with their hydrodynamical simulations. In this model we have two equations of the potentials from equations (\ref{eq:PDE_dU}) and (\ref{eq:PDE_dPhi}) as
	\begin{eqnarray}
	\label{eq:2potentials_dU}
		&&0= \DDr{R}\delta U \nonumber \\
		&&+ \left[\del{R}\ln\left(\frac{s}{L}\right)-\frac{2m\Omega}{sR}+\frac{1}{R}
			+\frac{\del{R}\rho_{_\mathrm{eq}}}{\rho_{_\mathrm{eq}}}+\frac{m\kappa^2}{2\Omega sR}\right]\Dr{R}\delta U \nonumber \\
	&& + \left[-\frac{L}{s}\del{R}\left(\frac{2m\Omega}{RL}\right)+\frac{L}{c_{_\mathrm{eq}}^2}
			-\frac{2m\Omega}{sR}\left(\frac{1}{R}+\frac{\del{R}\rho_{_\mathrm{eq}}}{\rho_{_\mathrm{eq}}}\right)-\frac{m^2}{R^2}\right]\delta U\nonumber \\
			&&- \frac{L}{c_{_\mathrm{eq}}^2}\delta\Phi,
	\end{eqnarray}
and
	\begin{equation}
			\label{eq:2potentials_dPhi}	
		\left(\DDr{R}+\frac{1}{R}\Dr{R}-\frac{m^2}{R^2}\right)\delta\Phi = 4\pi\rho_{_\mathrm{eq}}\left(\frac{d\rho}{dp}\right)_\mathrm{eq}(\delta U-\delta\Phi).
	\end{equation}
Here $\rho_{_\mathrm{eq}}$ is the equilibrium density, $\Omega$ is the angular velocity of equilibrium flow, $\kappa^2= 2\Omega\left(2\Omega + R\del{R}\Omega\right)$ is the epicyclic frequency squared, $s=\sigma-m\Omega$ is the frequency seen from a co-moving observer to the equilibrium flow, $L=s^2-\kappa^2$ and the sound speed of equilibrium fluid is $c_{_\mathrm{eq}}$. 
If $\Re[s]=\Re[\sigma]-m\Omega(R)=0$ at  a point $R$ inside a star, we call the corresponding solution to the linearised 
equations as `corotating' and a zero of $\Re[s]$ is called `corotation radius'. If $s$ is purely real, its zero is called `corotation singularity' of the perturbed equations since equation (\ref{eq:2potentials_dU}) becomes singular there. When a mode becomes growing or damping in time, the frequency
becomes complex and the zero of $s$ does. We still call this zero a corotation singularity, which is a simple pole
of equation (\ref{eq:2potentials_dU}) in the complex $R$-plane.

\subsection{Boundary conditions}
Since the equations (\ref{eq:2potentials_dU}) and (\ref{eq:2potentials_dPhi}) form
a single forth order ordinary differential equation, four conditions on $\delta U$ and $\delta\Phi$
are required. For a star (not a disc or torus), we impose regularity conditions of $\delta U$ and $\delta\Phi$
at the origin. This results in $\delta U\sim R^m$ and $\delta\Phi \sim R^m$ as $R\to 0$. One of the
other two conditions is the free boundary condition at the stellar surface, $\Delta h = 0$, where
$\Delta$ means taking Lagrangian perturbation to the variable that follows it. For a scalar variable
$h$ we have $\Delta h = \delta h + \xi^a\del{a}h$, where the vector $\xi^a$ is the Lagrangian displacement
vector that connects the location of unperturbed and perturbed fluid elements that are identified. By using perturbed velocity components the displacement vector is written as
	\begin{eqnarray}
		\xi^R &=& \frac{i}{s}\delta v^R,\\
		\xi^\varphi &=& \frac{i}{s}\delta v^\varphi-\frac{\del{R}\Omega}{s^2}\delta v^R
			-\frac{\del{z}\Omega}{s^2}\delta v^z,\\
		\xi^z &=& \frac{i}{s}\delta v^z,
	\end{eqnarray}
	by using $\partial_t\xi^a = \Delta v^a$ \citep{Friedman_Schutz1978a}.
Then the boundary condition in the cylindrical model reduces to
	\begin{equation}
		\label{eq:surfaceBC1}
		sL(\delta U-\delta\Phi) + s\del{R}h~\Dr{R}\delta U - \frac{2m\Omega\del{R}h}{R}\delta U = 0,
	\end{equation}
at $R=\Req$. The other condition for $\delta\Phi$ is that it is smoothly matched to the vacuum
gravitational potential at the surface,
	\begin{equation}
		\label{eq:surfaceBC2}
		\left.\Dr{R}\delta\Phi + \frac{m}{R}\delta\Phi\right|_{R=\Req} = 0.
	\end{equation}

\subsection{Equilibrium star}
Equilibrium models on which we perform linear perturbation analysis are obtained
by solving the system of equations above by Hachisu's self-consistent field (HSCF) method \citep{HSCF} on the stationary and axisymmetric assumptions. The method solves for physical variables as functions of coordinates in the meridional plane of a star. With the polytropic EOS, its angular velocity 
profile is a function of the distance from the rotational axis. We adopt the so-called `j-const.' law for the $\Omega$ profile, 
	\begin{equation}
		\Omega (R) = \frac{j_0}{R^2+d^2},
		\end{equation}
where $j_0$ and $d$ are constants and $R$ is the distance from the rotational axis (i.e., the cylindrical radial coordinate). As in \citet{Saijo_Yoshida2016},  we normalize the degree of differential rotation as $d=A \Req$, where $\Req$ is the equatorial radius of the star. Our numerical HSCF code \citep{Yoshida_Eriguchi1995} is tested to reproduce the original results of \citet{HSCF}. The equilibrium models are computed on spherical polar grid points and the numbers of grid points are $n_r=300$ in the radial direction and $n_\theta=100$ in the polar direction ($0\le \theta \le\pi/2$) inside a star. The number of spherical harmonics to compute gravitational potential is 26 (Legendre polynomials $P_{2\ell}(\cos\theta)$ of  $0\le \ell \le 25$ are taken into account.).

\begin{figure}
\includegraphics[width=\columnwidth]{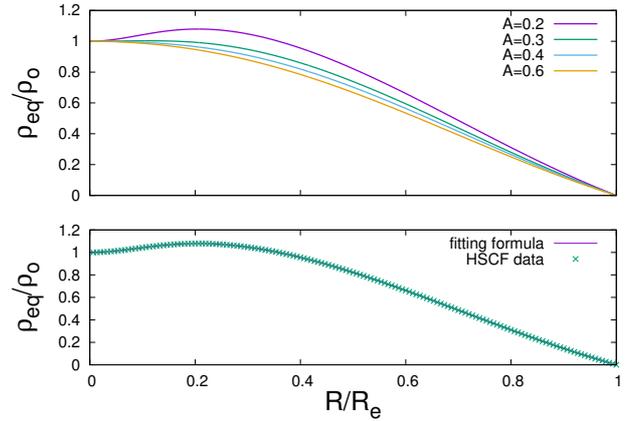}
\caption{Equilibrium density $\rho_{\mathrm{eq}}$ profile on the equatorial plane for the case with $T/W=3.95\times 10^{-2}$. Equilibrium data computed by HSCF scheme is fitted with equation (\ref{eq:equil density}). They are
normalized by the density at the origin $\rho_\mathrm{eq}$. Different value of degree of differential rotation $A$ is compared above. Below the data of $\rho_{_\mathrm{eq}}$ computed on a uniform grid points are compared with the corresponding fitting
function ($A=0.2$). In this case, the parameters in equation (\ref{eq:equil density}) are $n_1=29.54$, $n_2=-34.26$, 
$n_3=3.722$, $d_1=23.91$, $d_2=9.181$ and $d_3=7.498$.
}
\label{fig:equil rho}
\end{figure}

The equilibrium density profile on the equatorial plane is fitted by an analytic form
	\begin{equation}
		\frac{\rho_{_\mathrm{eq}}(R/\Req)}{\rho_\mathrm{o}} = \frac{1 + n_1 (R/\Req)^2 + n_2 (R/\Req)^4 + n_3 (R/\Req)^6}{1+  d_1 (R/\Req)^2 + d_2 (R/\Req)^4 + d_3 (R/\Req)^6},
	\label{eq:equil density}
	\end{equation}	
where $\rho_\mathrm{o}=\rho_{_\mathrm{eq}}(R/\Req=0)$, $n_i$'s and $d_i$'s are constants (notice that they depend on EOS, $A$ parameter, central density and $T/W$). Density profiles for different $A$ parameter are plotted in Figure \ref{fig:equil rho} for fixed $T/W=3.95\times 10^{-2}$. In this case, we have a star with
off-center density maximum when $A$ is less than $\sim 0.35$. The fitting is done by using {\tt fit} function of {\tt gnuplot}. Apart from stars with high degree of differential rotation ($A<0.2$),
asymptotic standard errors of our fitting are less than 1 \%.  For $A\ge 0.2$ residuals of some of those coefficients is a few tens of percent. The derivatives of equilibrium density with respect to $R$ are analytically computed from the fitting
formula. 

An equilibrium star is characterized by the degree of differential rotation $A$ and a ratio of rotational energy $T$ to gravitational binding energy $W$,
	\begin{equation}
		T = \frac{1}{2}\int \rho_{_\mathrm{eq}} R^2\Omega^2 dV, \qquad W = -\frac{1}{2}\int \rho_{_\mathrm{eq}}\Phi dV.
		\label{eq:T and W defined}
	\end{equation}

Hereafter we adopt the normalization such that $G=1$ and the coefficient of polytropic equation of state
$p=K\rho^\gamma$ to be unity, thus $K=1$. We fix $\gamma=2$.

\subsection{Numerical computation of eigenmodes}
To solve the two point boundary value problem, we adopt a shooting method in which the perturbed equations are solved from the origin and the surface of the star to a matching point $R=\Rmatch$ in the middle of the star. The integration
requires a careful treatment when there is a corotation radius. We first describe how to bridge the numerical solution at a corotation radius, then briefly explain how the matching and extraction of an eigenmode is done.
\subsubsection{Bridging the solution at a corotation radius\label{sec:bridging_explained}}
Consider a complex frequency $\sigma = \sigr + i\sigi$ in which $\sigr=\Re[\sigma]$ and $\sigi=\Im[\sigma]$. We say that a mode has corotation radius if $\sigr-m\Omega(R)=0$ has solutions $R=\Rc$ and its roots $\Rc$ are called corotation radii.  Equation (\ref{eq:2potentials_dU}) has a singular point on the real $R$-axis at $R=\Rc$ if the frequency is purely real ($\sigi=0$). 
For the numerical integration of the system of the differential equation we assume $\sigma$ is purely real ($\sigi=0$), In this case
we need to specify how we treat the singularity in our numerical integration on the $R$-axis. For a given trial frequency $\sigma$ we first check if there is a corotation radius inside the star (since our model has a monotonic profile of angular frequency $\Omega(R)$, we have at most one corotation radius.). If we find $\Rc$, the numerical
integration of equations (\ref{eq:2potentials_dU}) and (\ref{eq:2potentials_dPhi}) is performed to a radius close to the corotation, $R=\Rc-\epsC~(0<\varepsilon_\mathrm{c}\ll \Rc)$. Then we bridge the solution to $R=\Rc+\epsC$ where the bridged solution is used as the initial data for numerical integration from $\Rc+\epsC$ to $\Rmatch$. For the bridging at $R=\Rc-\epsC$, we match a numerical solution $(\delta U, \frac{d\delta U}{dR}, \delta\Phi, \frac{d\delta\Phi}{dR})$ to Frobenius series
solutions (appendix~\ref{appendix:corotation}) truncated at the lowest order. The series are then evaluated at $R=\Rc+\epsC$ which form the initial data of the numerical integration outside the corotation radius.

\begin{figure}
\includegraphics[width=\columnwidth]{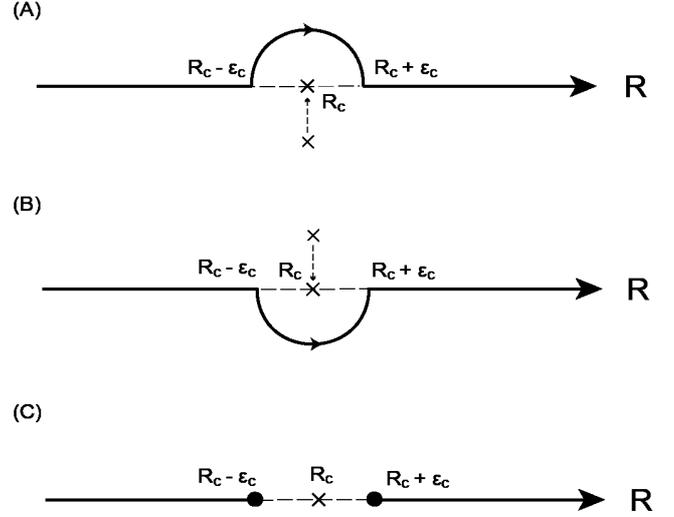}
\caption{Schematic figures of integration paths for different types of eigenmodes when they have a corotation radius. The numerical integration is done on the real $R$ axis. (A) For a dynamically unstable mode ($\sigi>0$) the corotation singularity resides beneath the real $R$ axis. In the limit of small imaginary part, the original integration path on the real $R$ axis (the dashed portion of the axis, $\Rc-\epsC \le \Rc \le \Rc+\epsC$) needs
to be modified so that the path of a semi-circle in the upper half plane goes around the singularity $R=\Rc$. (B) For dynamically stable mode with damping, the corotation radius is averted from below. (C) For purely real mode with a corotation radius, the numerical solution on the left of the corotation radius $R=\Rc-\epsC$ needs to be matched with another numerical solution on the right of the corotation $R=\Rc+\epsC$. The matching permits an extra parameter that cannot be fixed by a physical condition, which results in an emergence of the corotation spectrum of eigenfrequency (\citet{Watts_etal2003, Yoshida2006}). }
\label{fig:integralpath}
\end{figure}

The bridging of solutions is done in the following three ways (Figure \ref{fig:integralpath}) depending on the mode nature. 

(A) Unstable mode that grows in time. Near the corotation radius, $s=0$ gives
	\begin{eqnarray}
		0&=&\sigr + i\sigi - m(\Omega(\Rc) + \Omega'(\Rc)(R-\Rc) \nonumber \\
		&& + {\cal O}[|R-\Rc|^2]) \nonumber \\
		&=& i \sigi -m[ \Omega'(\Rc)(R-\Rc) + {\cal O}[|R-\Rc|^2]],
		\end{eqnarray}
where $\Omega'=\frac{d\Omega}{dR}$ and we expanded $\Omega$ around $\Rc$. Thus we have
	\begin{equation}
		\label{eq:approximate pole}
		R = \Rc - i \left(\frac{\sigi}{-m\Omega'(\Rc)}\right),
	\end{equation}
neglecting the higher order terms. We have $\Omega'<0$ and $\sigi>0$ for an unstable mode, thus the singularity as a simple pole of equation (\ref{eq:2potentials_dU}) resides in the lower half of the complex $R$ plane. Since in our numerical procedure the pole is on the real axis for $\sigi$ is assumed to be small compared to $\sigr$,
\footnote{
The imaginary part of the frequency can be perturbatively 
estimated only after we numerically integrate and match the solution. Therefore an (approximate) integration path should be specified before the numerical integration.}
we should indent the integration path by the semicircle of the radius $\epsC$ in the upper half plane
to avoid the singularity from $R=\Rc-\epsC$
to $R=\Rc+\epsC$.  
We develop a Frobenius series solution around $\Rc$ and use it for bridging with taking account of the change of the argument of $R$ in amount of $-\pi$ (from $r=R+\epsC \eNap^{i\pi}$ to $R+\epsC \eNap^{i\cdot 0}$). For our numerical results here we adopt $\epsC=10^{-4}R_\mathrm{e}$. For numerical integrations double precision {\tt LSODA} routine from {\tt ODEPACK} \citep{odepack} is used by fixing the tolerance of relative and absolute errors to be $10^{-8}$, automatically varying the order of accuracy of the solver. The routine automatically switches between non-stiff solver (Adams-Moulton method) and stiff solver (BDF (backward differential formula) method).

(B) Stable mode that damps in time. In this case we have a positive imaginary part of the pole in complex $R$-plane.
Thus the appropriate numerical path is indented below the pole on the real axis. The change of argument of $R$
in amount of $+\pi$ needs to be taken into account for the bridging, since $R$ on the path changes from $r=R+\epsC \eNap^{-i\pi}$ to $R+\epsC \eNap^{i\cdot 0}$.

(C) Purely real mode belonging to a continuous spectrum. The corotation radius is a regular singular point on the real $R$-axis and a Frobenius series
solution is obtained around the corotation radius \citep{NISTHandbook}. Its indicial equation has two solutions $0$ and $1$, thus one
of the series solutions is regular and the other has a logarithmic term. The most general solution in terms of distributions (generalized functions) has a discontinuity in the first derivative of $\delta U$ (see section \ref{sec: continuous spectrum} below). 
The step size is arbitrary in a sense that
no physical constraint fixes its value. They call a special case with zero discontinuity as `zero-step solutions'. Here we
compute the zero-step solutions as example of purely real modes belonging to corotation spectra except in Figure \ref{fig:continuous spectrum}. 

\subsubsection{Lindblad radius}
As in a gaseous disc around a star, a non-axisymmetric mode in a differentially rotating star may have 
Lindblad resonances at which $L=(\sigma-m\Omega)^2-\kappa^2=0$, where $\kappa^2=2\Omega\left(2\Omega+R\frac{d\Omega}{dR}\right)$ is the epicyclic frequency squared. The unstable modes with corotation radii in our models have outer Lindblad resonances $\ROL$ at which $\sigma-m\Omega(R_\mathrm{OL}) = \kappa(\ROL)$. Although equation (\ref{eq:2potentials_dU}) appears to be singular at the outer Lindblad resonance, a careful look at the behaviour of coefficients of the equation reveals that the resonance is a regular point of the differential equation and we have a regular solution of the differential equations there (see appendix~\ref{appendix:2potentials}).
When an outer Lindblad resonance appears, we numerically solve the perturbed equations to $R=\ROL-\varepsilon_\mathrm{OL}$ for $\epsOL>0$. Around $R=\ROL$ we have a regular series expansion of the solution
whose coefficients are determined by matching it with the numerical solution. The series expansion is then used
as the initial condition of numerical integration at $R=\ROL+\epsOL$. We typically use $\epsOL=10^{-4}\Req$ and the series expansion is truncated up to the ${\cal O}[\epsOL^3]$ terms.

\subsubsection{Numerical matching to find eigenmodes}
To obtain an eigenmode we first take two linearly independent solutions at the origin that satisfy the regularity conditions there. We then integrate equations (\ref{eq:2potentials_dU}) and (\ref{eq:2potentials_dPhi}) numerically on the real $R$-axis to an arbitrary matching point $R=\Rmatch$ by fixing a real $\sigma$. For our numerical computation $\Rmatch$ is equal to $\Req/2$  when there is no corotation radius, and to $(\ROL+\Req)/2$ when there is a corotation radius.
If $\sigma$ satisfies the corotation condition $\sigma-m\Omega=0$ between the origin and $R=\Rmatch$, we 
follow the procedure described in section \ref{sec:bridging_explained}.
At the matching point $R=\Rmatch$, we have a general solution that satisfies the two boundary conditions,
$\boldsymbol{y}_L = A^{(1)}\boldsymbol{y}_1 + A^{(2)}\boldsymbol{y}_2$, where $\boldsymbol{y}_i = {}^T(\delta U, \frac{d\delta U}{dR}, \delta\Phi, \frac{d\delta\Phi}{dR})_i$ for $i=1,2$. Here $A^{(i)}$ ($i=1,2$) are complex constants and they corresponds to two linearly independent solutions that satisfy the boundary condition at the origin. Next we take two linearly independent solutions that satisfy the boundary conditions at the surface (equations (\ref{eq:surfaceBC1}) and (\ref{eq:surfaceBC2})) and numerically integrate equations (\ref{eq:2potentials_dU}) and (\ref{eq:2potentials_dPhi}) on the real $R$-axis to a matching point $R=\Rmatch$. We obtain $\boldsymbol{y}_R = A^{(3)}\boldsymbol{y}_3 + A^{(4)}\boldsymbol{y}_4$ where $\boldsymbol{y}_i ~(i=3,4)$ are two linearly independent solutions that satisfy the surface boundary conditions. The matching is done by equating $\boldsymbol{y}_L$ and $\boldsymbol{y}_R$ at $R=\Rmatch$,
	\begin{equation}
		\left[\boldsymbol{y}_1~~\boldsymbol{y}_2~~-\boldsymbol{y}_3~~-\boldsymbol{y}_4\right]\left(\begin{array}{c}A^{(1)}\\ A^{(2)}\\ A^{(3)}\\ A^{(4)}\end{array}\right) = \boldsymbol{0},
	\label{eq:matching}
	\end{equation}
whose $4\times 4$ coefficient matrix $\left[\boldsymbol{y}_1~~\boldsymbol{y}_2~~-\boldsymbol{y}_3~~-\boldsymbol{y}_4\right]$ is a function of $\sigma$.
A zero of determinant of the coefficient matrix of equation (\ref{eq:matching}) gives eigenfrequency which may be complex in general. An imaginary part of $\sigma$, if existed, is assumed to be small compared to the real part and is solved for as a small perturbation to the real part. Thus the numerical integrations are done by assuming $\sigma$ is real. On the other hand $\delta U$ and $\delta\Phi$ are allowed to have complex value in the integration because of the possible appearance of a corotation radius.
We compute the modulus of the determinant ${\cal W}$ on the real $\sigma$ axis of the complex $\sigma$-plane. If we have a real zero of ${\cal W}(\sigma)$, it is an eigenmode with a purely real eigenfrequency. Assuming ${\cal W}(\sigma)$ is an analytic function near the real $\sigma$ axis in the complex $\sigma$-plane we also find an eigenmode with a complex eigenfrequency if the eigenfrequency lies close to the real $\sigma$-axis. We identify a local minimum of ${\cal W}(\sigma)$ as a function of real $\sigma$ and fitting the function around the minimum by a parabola, ${\cal W}(\sigma)=\alpha_{_0}\sigma^2 + \alpha_{_1}\sigma +\alpha_{_2}$ where $\alpha_{_0}, \alpha_{_1}, \alpha_{_2}$ here are real numbers. The complex solution of the quadratic equation of the parabola is an approximate complex eigenfrequency. We use a method of trisection for the root-findings of ${\cal W}$. 

\section{Results}
\label{sec: Results}
\subsection{Eigenmode spectrum}
We briefly overview the structure of eigenmode spectrum of our model.
Here we call the region (band) of frequency for which we may find a corotation radius inside a star as a corotation band of frequency or corotation band. Inside the corotation band, we find both complex (growing/damping) modes as well as purely real
modes. Purely real modes belong to van Kampen
type peculiar modes with continuous spectrum
of eigenfrequency \citep{VanKampen1955, Stix_waves_in_plasmas}.

\subsubsection{Purely real continuous eigenmodes in the corotation band of frequency}
\label{sec: continuous spectrum}
In the corotation band of frequency, we generally find an infinite number of purely real
frequency modes and possibly find modes with complex eigenfrequency. 
The former belongs to the so-called continuous spectrum of frequency, since
the eigenfrequency that satisfies the physical boundary conditions are dense in a subset of real $\sigma$ and spanning  certain range of frequency domain inside the corotation band.

The continuous spectrum of modes is understood as van Kampen mode. The linearised equation
for $\delta U$ is cast into a form in which the corotation factor $(\sigma-m\Omega)$ is multiplied to the
highest order derivative term and all the other terms are regular. When we divide the both hand side of the
equation by the corotation factor, we formally have a delta function term as $C~\delta(\sigma-m\Omega)$ where
$C$ is a real constant, if $\sigma$ is purely real \citep{Dirac1947, VanKampen1955}. The constant is not
determined by physical arguments, but we may choose a value of it so that the boundary conditions at the
origin and the surface is satisfied for any real $\sigma$. This is the origin of the continuous spectrum
of eigenfrequency. When $\sigma$ is not purely
real, the delta function term does not arise and only discrete eigenfrequency satisfies all the boundary
conditions simultaneously. This situation can be explained by the matching of the Frobenius series solutions
on both sides of the corotation radius \citep{Watts_etal2003}. The corotation radius is a regular singular point of the equation for $\delta U$, whose series solution has an extra constant that is not fixed by the boundary conditions nor any physical argument at the corotation.
\footnote{The situation may be different when we introduce small viscosity to the fluid. Even a small viscosity
drastically changes the nature of the solution by forming a boundary layer close to the corotation radius \citep{Lockitch_etal2004}. Viscosity regularize the singularity and the extra degree of freedom disappears.}
 Due to this degree of freedom of the constant,  we may satisfy all the boundary conditions for any
real frequency. The constant amounts to discontinuity of gradient of $\delta U$ on both side of 
the corotation radius, thus the eigenfunction profile of $\delta U$ is continuous but have a kink there.

\begin{figure}
\includegraphics[width=\columnwidth]{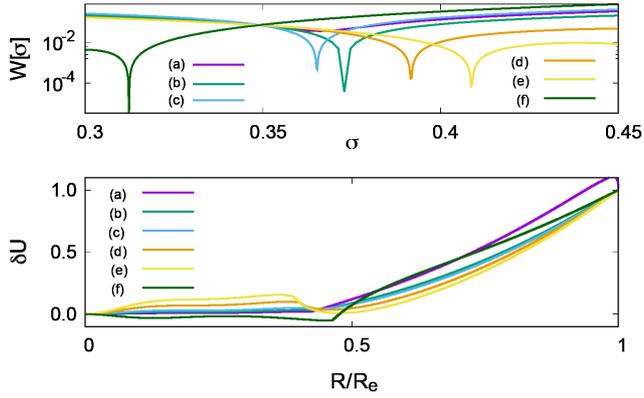}
\caption{Top: Determinant ${\cal W}$ of the coefficient matrix of equation (\ref{eq:matching}) as a function of real frequency $\sigma$. Frequency $\sigma$ is normalized by $\OmegaO$, the rotational frequency of equilibrium star at the origin. Solid line (a) is for a complex mode case with the integration path of (A) of section \ref{sec:bridging_explained}. Curves (b)-(f) correspond to different values of discontinuity in $\frac{d\delta U}{dR}$ at the corotation radius; (b): zero step solution \citep{Watts_etal2005}; (c), (d), (e), (f): discontinuity parameter $\chi=-1000, 800, 1000, 4800$, respectively.
Each trough corresponds to eigenfrequency. Equilibrium parameters are $A=0.2$ and $T/W=3.95\times 10^{-2}$. Bottom: Eigenfunctions $\delta U$ for the eigenmodes
whose eigenfrequencies are defined as the roots of ${\cal W}[\sigma]$. For (a) the real part of $\delta U$ is plotted as a function of $R/\Req$.
}
\label{fig:continuous spectrum}
\end{figure}

We see purely real frequency modes in the corotation band of frequency. These modes seems
discrete in the corotation band when we fix how we match the solutions on the real axis $R$ around the corotation radius, i.e., how we choose the size of discontinuous jump of the first radial derivative of $\delta U$. 
In Figure \ref{fig:continuous spectrum} such purely real modes are displayed. These solutions all belongs to generalized $f$-mode in a continuous spectrum. We parametrize discontinuity
as follows. As seen in appendix~\ref{appendix:corotation}, a series solution around the corotation radius $\Rc$
is written as a linear combinations of two solutions. One of the solution $y_1(R)$ is regular at $R=\Rc$.
The other solution $y_2(R)$ is written as
	\begin{equation}
		y_2(R)= b_{_0} + b_{_1}(R-\Rc)^2 + b_{_2} (R-\Rc)^2 + \cdots + y_1(x)\log|R-\Rc|,
	\end{equation}
	where $b_{_i}$ are constants. We introduce a parameter $\chi$ so that $b_1$ is replaced with $b_1(1+\chi H(x))$, where $H(x)$ is Heaviside's step function. The extra factor $(1+\chi H(x))$ corresponds to the discontinuity in $\del{R}\delta U$ at $R=\Rc$. The case with $\chi=0$ is a zero step 
solution \citep{Watts_etal2005}. We see discrete zeroes of 
determinant ${\cal W}[\sigma]$ (see section \ref{sec:bridging_explained}) when we fix the discontinuous derivative
of $\delta U$ at the corotation radius. The zero of ${\cal W}[\sigma]$ changes its location continuously
when we change the matching of $d\delta U/dR$ in a continuous manner, thus forms a continuous 
spectrum \citep{Yoshida2006}.

\begin{figure}
\includegraphics[width=\columnwidth]{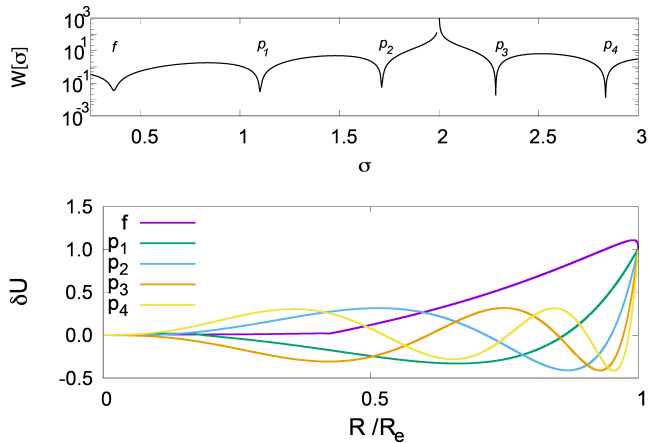}
\caption{Top:  ${\cal W}$ as a function of real frequency $\sigma$. Equilibrium parameters are $A=0.2$ and $T/W=3.95\times 10^{-2}$. For $f$, $p_1$ and $p_2$-modes inside the corotation band, the integration path (A) is taken. $\sigma=2$ is the higher frequency edge of the corotation band. 
The eigenfrequencies of modes are; $f: 0.3662+1.4152\times 10^{-2} i$, $p_1: 1.1013+7.8206\times 10^{-4} i$, $p_2: 1.7126+5.8305\times 10^{-5} i$ for unstable modes, and $p_3: 2.2841$, $p_4: 2.8358$ for purely real modes outside the corotation band. Frequency $\sigma$ is normalized by the rotational angular frequency of equilibrium star at the origin. Bottom: Eigenfunction $\delta U$ profile for the first five eigenmodes.
}
\label{fig:A=0.2 T/W=3.95e-2 eigenmodes}
\end{figure}
We plot ${\cal W}[\sigma]$ and the eigenfunctions for unstable modes inside the
corotation band and the purely real modes outside the corotation band
in Figure \ref{fig:A=0.2 T/W=3.95e-2 eigenmodes}.  Notice that we choose the same equilibrium model as
Figure \ref{fig:continuous spectrum} for comparison.
The features found here are essentially the same as those in \citet{Saijo_Yoshida2016}. Here $f$-, $p_1$- and $p_2$-modes are computed as a complex (growing) mode while $p_3$- and $p_4$-modes are purely real discrete modes outside the corotation band. The corotation band is defined as $\Omega(\Req)/\OmegaO\le \sigma/\OmegaO \le m$, where $\Omega(\Req)=7.69\times 10^{-2}\OmegaO$ and $m=2$ for the model shown here. Outside the corotation band, there is no continuous spectrum.
It is noticed that the eigenfunctions nicely follows the rule of
stellar pulsation with corotation, i.e., the number of nodes between
the corotation radius (or the centre if there is no corotation radius
inside the star) and the surface classifies the mode \citep{Saijo_Yoshida2016}.

\subsubsection{Complex modes - growing and damping}
\begin{figure}
\includegraphics[width=\columnwidth]{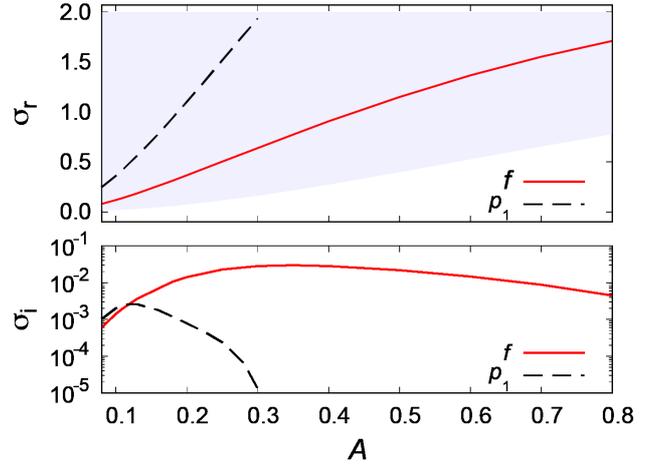}
\caption{
Frequencies of growing $m=2$ $f$- and $p_1$-modes as a function
of the parameter $A$ of degree of differential rotation.
We fix $T/W=3.95\times 10^{-2}$ here.
Frequency $\sigma$ is normalized by $\OmegaO$, the rotational frequency of equilibrium star at the origin. In the upper panel the real part of $\sigma$ for $f$-mode (solid) and $p_1$-mode (dashed) are plotted. The shaded region is the corotation band of frequency. The upper bound of the corotation frequency is $\sigr=2$ with our normalization. Thus the mode shown here is inside the corotation band of spectrum. In the lower panel imaginary part of $\sigma$ is plotted.}
\label{fig:TW3.95e-2 sigma(A)}
\end{figure}

\citet{Watts_etal2005} conjectured that a $f$-mode becomes dynamically unstable
when it is inside the corotation band of spectrum. \citet{Passamonti_Andersson2015} showed that, 1) the $f$-mode
becomes unstable when it enters the corotation band, and 2) the growth rate of the mode inside the corotation band increases as the mode frequency is farther away from boundaries of the corotation band. As for 1) it is the case by construction in our model, because we may choose the path around the corotation radius in such a way that the mode grows exponentially. If the sound wave trapping conditions (the regularity at the origin and the free boundary condition at the stellar surface) are all satisfied and if the characteristic frequency of the wave admits
an appearance of corotation radius in the star, we always have a mode that is defined by the integration path (A) in
section \ref{sec:bridging_explained}. The same holds for damping  modes. Actually the damping mode appears as the complex conjugate solution to the growing mode.
To see the point 2) above we plot the complex frequency of unstable $f$-mode as a function of $A$ which parametrizes the degree
of differential rotation in Figure \ref{fig:TW3.95e-2 sigma(A)}. We fix $T/W$ of the equilibrium stars to see the effect
of differential rotation. As we see in the figure, the imaginary part of the mode frequency increases as its distance from the edges of the corotation band of frequency becomes larger. This is consistent with the results obtained by \citet{Passamonti_Andersson2015}, who argue that the growth rate of the instability should depend on the position of the mode relative to the boundary of the corotation region, being more rapid deep inside the corotation region and vanish at the edges of the corotation band. We also note that $p_1$-mode inside the corotation band is unstable and shares the same characteristics as $f$-mode, i.e., the imaginary part of the frequency is maximized in the middle of the corotation band and vanishes at the edges of the corotation band.

\subsection{Growing eigenmodes - nature of the instability}
Hereafter we mainly focus our attention on growing eigenmodes which have a corotation radius. 

\subsubsection{Diagnosing by canonical angular momentum and its flux distribution}
Here we compute the density of canonical angular momentum of an unstable eigenmode defined in \citet{Friedman_Schutz1978a} and its flux derived by using the expression there. Canonical angular
momentum is defined as a quadratic form of Lagrangian displacement vector of perturbation, which
is conserved when the displacement vector satisfies the linearly perturbed equation to axisymmetric
fluid configuration. For the dynamically unstable perturbation, the value of canonical angular momentum
vanishes.

Canonical angular momentum density, which is an integrand of canonical angular momentum, is first applied to characterize dynamical instability of rotating stars
by \citet{Saijo_Yoshida2006}. They show if an unstable mode has a corotation radius the angular momentum density abruptly change its sign at the corotation radius. By introducing a concept of canonical angular momentum flux, we here see more clearly what is driving the instability in these unstable modes with corotation radius.

As is seen in appendix~\ref{sec:jc and flux}, canonical angular momentum density $\jc$ and its flux vector $f^i$ satisfies the following conservation equation as 
	\begin{equation}
		\del{t}\jc + \nabla_j f^j = 0.
		\label{eq:jc conservation}
		\end{equation}

\begin{figure}
\includegraphics[width=\columnwidth]{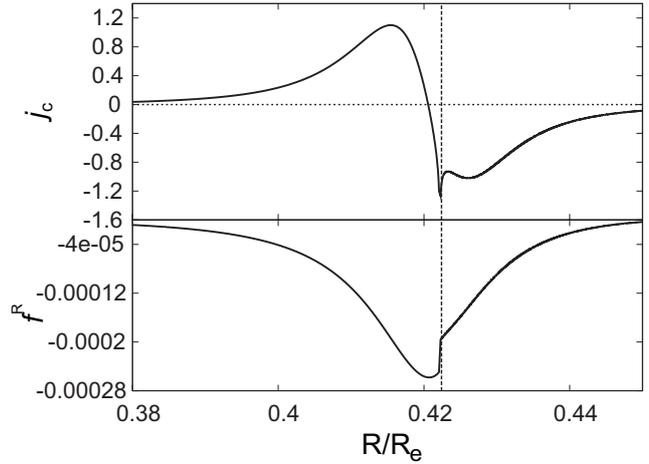}
\caption{Canonical angular momentum density and its flux distribution of $m=2$ $f$-mode as 
a function of radial coordinate.
In the upper panel is shown the integrand $\jc$ of canonical angular momentum as a function of the stellar
radius. Corresponding flux $f^R$ is plotted in the lower panel. The vertical dashed line marks the location of
the corotation radius. The equilibrium star has $T/W=3.95\times 10^{-2}$ and $A=0.2$. The eigenmode is $f$-mode in Figure \ref{fig:A=0.2 T/W=3.95e-2 eigenmodes}.}
\label{fig:jc and flux m2f}
\end{figure}

\begin{figure}
\includegraphics[width=\columnwidth]{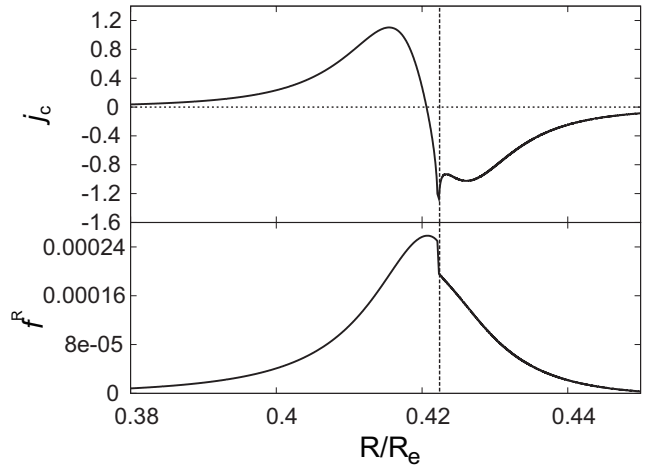}
\caption{The same as Figure \ref{fig:jc and flux m2f} except that the mode is damping (case (B) in section \ref{sec:bridging_explained}). The eigenmode is the complex conjugate mode of Figure  \ref{fig:jc and flux m2f}.}
\label{fig:jc and flux m2f damping}
\end{figure}

\begin{figure}
\includegraphics[width=\columnwidth]{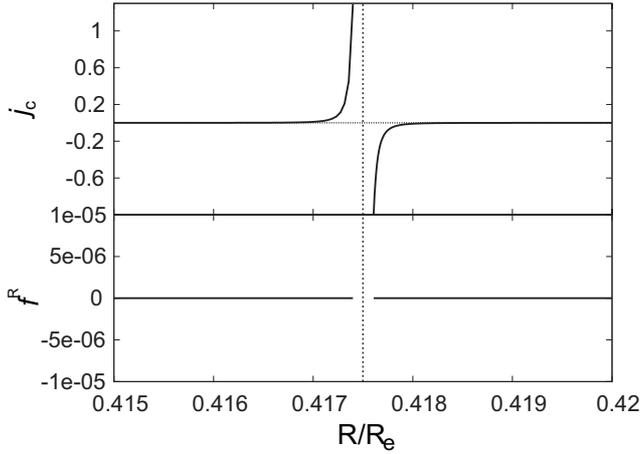}
\caption{The same as in Figure \ref{fig:jc and flux m2f} for $m=2$ $f$-mode near its corotation radius, except that 
the mode belongs to the continuous spectrum (corresponding to (b) of Figure \ref{fig:continuous spectrum}). The gap
of the curves around the corotation radius appears because we bridge the solution on the left to the right
at a finite distance from the corotation.}
\label{fig:jc and flux m2fkpath0}
\end{figure}

We plot $\jc$ and $f^R$ of $m=2$ $f$-mode around the corotation radius in Figure \ref{fig:jc and flux m2f}.  
The mode has a positive canonical angular momentum density on the left hand side of the corotation
radius. On the right side of the corotation radius the density is negative. This is a generic behaviour
of $\jc$ for modes having corotation radius (see also \citet{Saijo_Yoshida2006}). The flux has a negative
gradient on the left and a positive gradient on the right of the corotation radius.  According to equation (\ref{eq:jc conservation}) negative $\jc$ decreases as time passes if $\del{R}(Rf^R)$ is positive and as a result the absolute value of $\jc$ grows. On the other hand positive $\jc$ grows if $\del{R}(Rf^R)$ is negative. This is the case we see
in Figure \ref{fig:jc and flux m2f}. The growth of amplitude of $\jc$ means the perturbed variables grows.
It should be remarked that the integrated canonical angular momentum is conserved even though amplitude
of $\jc$ grows. In fact it is always zero. This can be interpreted as the corotation radius supplying canonical
angular momentum of opposite signs to drive the instability. For the complex conjugate mode of the growing
$f$-mode the flux has the opposite sign (Figure \ref{fig:jc and flux m2f damping}). That means the oscillation is
damped due to an action of corotation radius.
The situation is totally different for purely real modes inside the corotation band, seen in Figure  \ref{fig:jc and flux m2fkpath0}. For purely real mode, canonical angular momentum density diverges at the corotation radius. Its sign is positive on the left and negative on the right
of the corotation radius, as in Figure \ref{fig:jc and flux m2f}. Its flux is, however, zero everywhere. Therefore
the driving mechanism of the instability does not work and the mode is purely oscillatory.

\subsubsection{Wave-energy flux}
\citet{Drury1985} introduced an energy flux of wave motion inside a differentially rotating cylindrical fluid mass,
as a quadratic form of linear perturbation variables. The expression is motivated by the rate of `$PdV$- work'
and defined as $\Re[\delta p^* \delta v^R]$ where $\delta p$ and $\delta v^R$ are perturbed pressure
and radial velocity component. At the cylindrical surface of radius $R$, the fluid inside the surface
injects this amount of power per area, averaged over the phase of oscillation. We take into account
of the gravitational perturbation by replacing 
$\delta p$ with $\rho\delta U = \delta p + \rho\delta\Phi$ and define the energy flux $F_E$ as
	\begin{equation}
		F_E = \Re[\rho\delta U^* \delta v^R].
	\end{equation}
\begin{figure}
\includegraphics[width=\columnwidth]{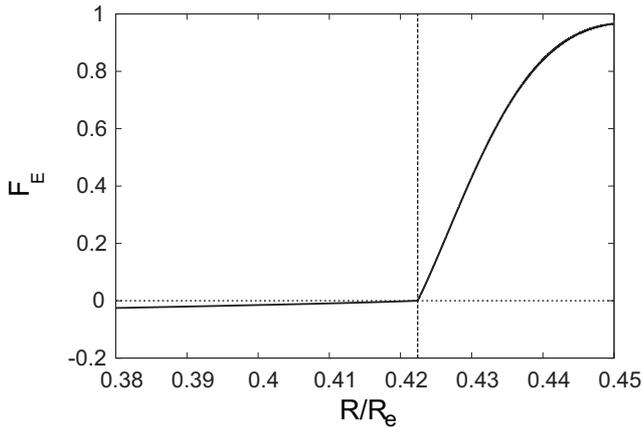}
\caption{Energy flux $F_E$ for $m=2$ $f$-mode with corotation radius. As in Figure \ref{fig:jc and flux m2f}, the dashed vertical line marks the location of the corotation radius.}
\label{fig:energy flux m2f}
\end{figure}
In Figure \ref{fig:energy flux m2f}, $F_E$ of unstable $m=2$ $f$-mode having corotation radius
is plotted as a function of radius near the corotation radius. We see the change of its sign at the
corotation radius, with energy flux pointing inward for the left hand side of the corotation radius
and outward for the right hand side of it. Drury argued this is the case of wave over-reflection 
at the corotation radius. The wave incident from the region inside the corotation radius is reflected back
with a larger amplitude, while the wave from outside to the corotation radius is also reflected
back with larger amplitude. We see the wave over-reflection may work at the corotation radius
and the corresponding trapped mode may be dynamically unstable.

On the other hand, the purely real mode that has a corotation radius shows $F_E=0$. This is also
the case with purely real eigenmodes that have no corotation radius (the mode outside the corotation band). Even if a mode has a corotation radius (that is, a mode that belongs to van Kampen mode), it is not necessarily dynamically unstable.

\subsubsection{Reflection coefficient of sound wave and growth timescale of unstable modes\label{sec:wave-overreflection}}
We next define a reflection coefficient of sound wave incident to corotation radius and estimate the growth rates
of the sound wave trapped inside the stars. These are compared with the imaginary parts of the eigenfrequencies of unstable modes.
We rewrite equation (\ref{eq:2potentials_dU}) in a stationary wave propagation equation as
\begin{equation}
\frac{d^2\eta}{d^2R}+ K^2\eta = S_{\eta},
\label{eq:local wave eq}
\end{equation}
where
\begin{eqnarray}
\eta &=& \delta U \eNap^{\frac{1}{2}\int \psi dR},\\
\psi&=& \del{R}\ln\left(\frac{s}{L}\right)-\frac{2m\Omega}{sR}+\frac{1}{R}
+\frac{\del{R}\rho_{_\mathrm{eq}}}{\rho_{_\mathrm{eq}}}+\frac{m\kappa^2}{2\Omega sR},\\
K^2 &=& \frac{L}{c_{s,0}^2}-\frac{m^2}{R^2}+\frac{2m\Omega}{sR}
	\left(\frac{\del{R}L}{L}-\frac{\del{R}\rho}{\rho}-\frac{\del{R}\Omega}{\Omega}\right)\nonumber\\
		&+& 
\left[R^2L^2(\del{R}\rho)^2
           -2R\rho L(-R\rho\del{R}L+L(\del{R}\rho+R\del{R}^2\rho))\right.\nonumber\\
           &+&\left.\rho^2(L^2-3R^2(\del{R}L)^2+2R L(\del{R}L+R\del{R}^2L))\right]\nonumber\\
           &&/(4R^2\rho^2 L^2),\\
S_{\eta} &=& \frac{L}{c_{_\mathrm{eq}}^2}\delta\Phi \eNap^{\frac{1}{2}\int \psi dR}.
\end{eqnarray}

\begin{figure}
\includegraphics[width=\columnwidth]{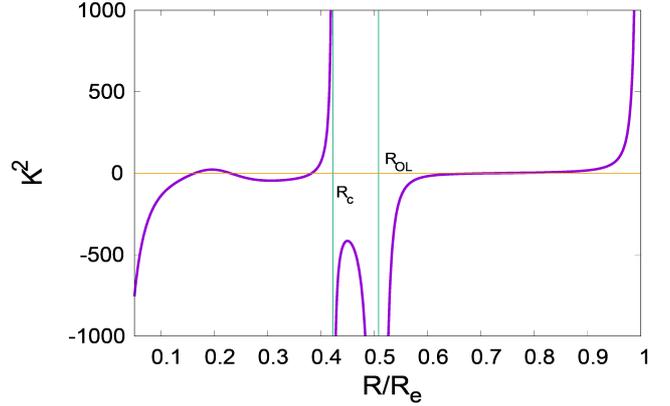}
\caption{$K^2$ of equation (\ref{eq:local wave eq}) for $m=2$ unstable $f$-mode. The equilibrium parameters are $A=0.2$ and $T/W =3.95\times 10^{-2}$. The mode has a corotation radius at $\Rc/\Req=0.4225$ and an outer Lindblad radius at $\ROL/\Req=0.5084$.
}
\label{fig:K2 wavenumber}
\end{figure}

\begin{figure}
\includegraphics[width=\columnwidth]{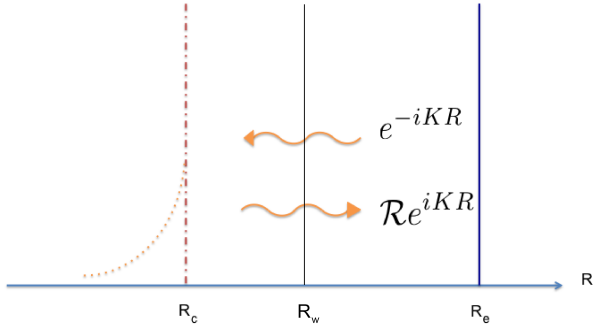}
\caption{Schematic picture of wave propagation and reflection at the corotation radius. Between the corotation radius $\Rc$ and the surface of the star $\Req$ the wave bounces back and forth to form the trapped wave. If the reflection coefficient ${\cal R}$ extracted at $\Rw$ is such that the modulus of complex amplitude exceeds unity, the wave is over-reflected at the corotation radius. }
\label{fig:schematic wave}
\end{figure}

\begin{figure}
\includegraphics[width=\columnwidth]{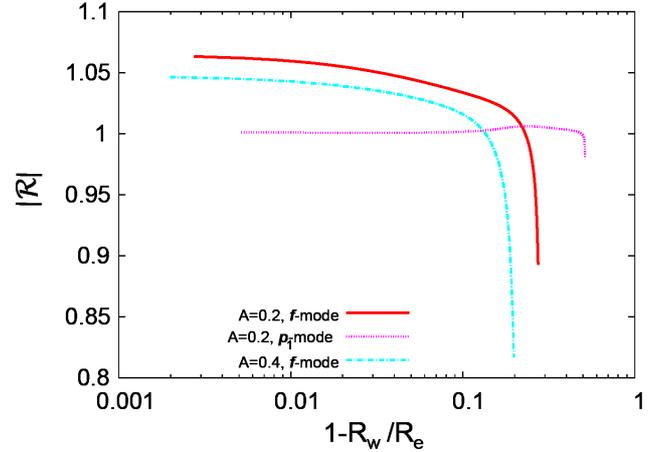}
\caption{Examples of reflection coefficients extracted at $R=\Rw$ as a function of $1-\Rw/\Req$. We fix $T/W=3.95\times 10^{-2}$.
}
\label{fig:Reflection coeffs}
\end{figure}

In Figure \ref{fig:K2 wavenumber}, we plot a typical profile of effective wave number squared $K^2$ in equation (\ref{eq:local wave eq}) for $m=2$ $f$-mode. Apart from a region around the stellar surface and the corotation radius $\Rc$, $K^2$ is mostly negative, which means a wave is evanescent in the high wave number limit (JWKB limit). 
For $K^2>0$, when $K^{-1}$ and $|d\ln(K^2)/dR|^{-1}$ are smaller compared to scale heights of equilibrium quantities, equation (\ref{eq:local wave eq}) is an equation of wave propagation of constant wave number $K$ with a source term. Moreover if the source term is small compared to other terms, a local solution for this equation
is written as
	\begin{equation}
	\eta = C_1 \eNap^{iKR} + C_2 \eNap^{-iKR},
		\label{eq:local wave}
		\end{equation}
with $C_1, C_2$ being constants. With recovering a time-dependent 
factor $\eNap^{-i\sigma t}$ we see the term with $C_1$
($C_2$) is a plane wave propagating outwards (inwards).  We divide a star into an inner and outer
part at a radius $R=\Rw$ (Figure \ref{fig:schematic wave}) whose outer domain satisfies $K^2>0$. Suppose we decompose our numerical solution $\eta$ in the form of equation (\ref{eq:local wave}) to determine the complex coefficients $C_1$ and $C_2$ in the outer domain. We may interpret the decomposition of the solution as follows. 
The wave solution (equation \ref{eq:local wave}) in $R>\Rw$ is regarded as a superposition of components travelling outward and inward. We suppose initiating a sound wave near the surface of the star that travels inward. The wave is reflected back around $\Rc<R<\Rw$. Thus a complex reflection coefficient ${\cal R}$ of the wave is defined as ${\cal R}=C_1/C_2$. When $|{\cal R}|>1$ the wave is over-reflected by the region $R<\Rw$.
The over-reflection of wave increases the amplitude of the wave by the factor of $|{\cal R}|$ in the characteristic period of $2\pi/\sigr$, during which the wave travels back and forth once in the star.
Thus the growth rate of trapped eigenmode $\hat{\sigma}_\mathrm{i}$ is estimated as
\begin{equation}
	\hat{\sigma}_\mathrm{i} = \frac{\sigr}{2\pi}\ln |{\cal R}|,
	\label{eq:sigi by |R|}
\end{equation}
where $\sigr$ is the frequency of the wave. This is directly compared with the imaginary part $\sigi$ of an unstable eigenmode. We evaluate the coefficient ${\cal R}$ by
	\begin{equation}
		{\cal R} = \frac{\frac{d\eta}{dR}+iK\eta}{\frac{d\eta}{dR}-iK\eta},
	\end{equation}
at points close to the surface (Figure \ref{fig:Reflection coeffs}). The numerical value of $|{\cal R}|$ is a function of $\Rw$ and we extrapolate them to the surface $R=\Req-0$ to determine $|{\cal R}|$. 

\begin{figure}
\includegraphics[width=\columnwidth]{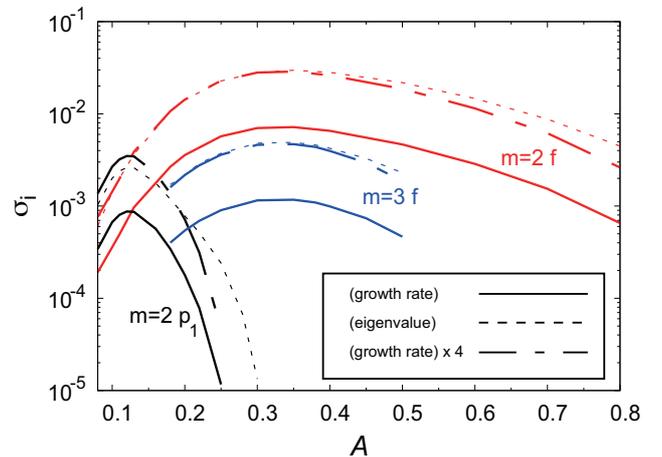}
\caption{The growth rates of $m=2$ and $m=3$ unstable modes computed by the reflection coefficients $|{\cal R}|$ are compared with the imaginary parts of complex eigenmodes $\sigma$. The equilibrium model has $T/W=3.95\times 10^{-2}$. The imaginary parts of the eigenmodes are plotted as dotted lines while the growth rates $\hat{\sigma}_\mathrm{i}$ for $f$- and $p_1$-modes computed with the reflection coefficients are plotted in thick solid curves. For comparison the growth rates multiplied by a factor of four are also displayed (dash-dotted). }
\label{fig:sigi by |R|}
\end{figure}

In Figure \ref{fig:sigi by |R|}, we compare the imaginary parts of unstable mode eigenfrequencies
with the growth rate computed by equation (\ref{eq:sigi by |R|}) for fixed $T/W$ models.
The growth rate computed by equation (\ref{eq:sigi by |R|}) are shown in thick solid curves, while the dashed curves are for the imaginary parts of $\sigma$ computed by the eigenmode analysis (the curves for $m=2$ $f$- and $p_1$-modes are the same ones as in the panel below of Figure \ref{fig:TW3.95e-2 sigma(A)}). 
Apart from the difference of a
factor $\sim 4$, the overall characteristics of growth rates $\tilde{\sigma}_\mathrm{i}$ are quite similar to the imaginary part of the eigenvalue. It suggests that the physical mechanism of the instability may be explained by the over-reflection of wave at a corotation radius and its trapping between corotation radius and the stellar surface.
As is seen in the figure, $m=3$ $f$-mode also shows the same trend.
We think the differences by the factor of four for $m=2, 3$ $f$-modes and $p_1$-modes is only a coincidence. The discrepancy of equation (\ref{eq:sigi by |R|}) with the true imaginary part of the eigenmode would come from the fact that the extraction of incoming and outgoing wave
as equation (\ref{eq:local wave}) works qualitatively well, though it is not accurate enough quantitatively.

We should make a final remark that the $f$-mode and low order $p$-modes we
are interested in have much smaller wave numbers and the physical picture of the sound wave travelling in the
propagation region ($K^2>0$ in Figure \ref{fig:K2 wavenumber}) may not be accurate. Propagation and evanescent region of sound wave may not be well-defined for these modes. We, however, expect that an important factor of a physical mechanism of the instability is grasped
in view of the qualitative agreement of growth rates with those solved directly by the eigenvalue problem.

\section{A possible mechanism of the low $T/W$ dynamical instability}
\label{sec: Discussions}
The natures of the instability of modes which has a corotation radius are compared with those of known
instability mechanisms of rotating stars or accretion discs. Firstly Papaloizou-Pringle instability \citep{Papaloizou_Pringle1984} may occur in highly differentially rotating system. The instability, however, is present when two surface waves interacts to exchange angular momentum \citep{Goldreich_etal1986_papI}.
Our stars do not have two surface waves to result in the instability, since they do not have an inner surface unlike in an accretion torus. 
%
Secondly a shearing flow may exhibit the well-known Kelvin-Helmholtz or more general barotropic instability \citep{Vallis2006book}.
Barotropic instability may be interpreted as an interaction of waves with positive and negative energy \citep{Cairns1979} and is
suggestive to our results of canonical angular momentum and flux distribution. 
The nature of the instability is, however, local and the modes with a higher wave number in azimuthal direction (larger $m$) tends to grow faster in barotropic instability. 
Such feature has not been found in low $T/W$ dynamical instability in
terms of hydrodynamic simulations. Spiral mode ($m=1$) or bar mode
($m=2$) plays a dominant role throughout the evolution (e.g., \citet{Saijo_etal2003}).
Therefore it is unlikely that the low $T/W$ instability is a manifestation
of barotropic instability.
%
%
%
Thirdly, as we spin the equilibrium star up,  a couple of modes may merge to form a complex conjugate pair of modes one of which is dynamically unstable. The merger occurs irrespective of the corotation of modes
and moreover a critical $T/W$ parameter of instability is ${\cal O}[0.1]$ \citep{Karino_Eriguchi2003}. This is the mechanism of classical bar-mode instability but may not be relevant here.
%
\begin{figure}
\includegraphics[width=\columnwidth]{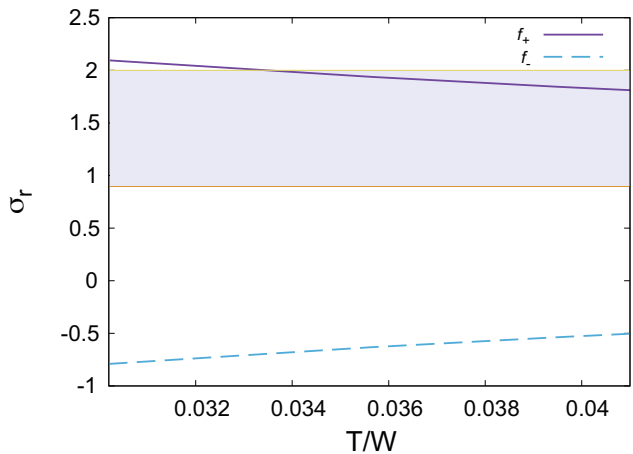}
\caption{Real part of eigenfrequencies of $m=2$ $f$-modes with prograde ($f_+$) and retrograde ($f_-$) pattern speed are plotted as a function of $T/W$. The degree of differential rotation is $A=0.9$. The prograde $f_+$ mode is the $f$-mode susceptible
to low $T/W$ instability.  It enters the corotation band (shaded area) at $T/W=0.0337$, in which the eigenfrequency becomes complex. Apparently Both of $f$-mode do not merge in the corotation band. }
\label{fig:f-plus and f-minus for A=0.9}
\end{figure}
In Figure \ref{fig:f-plus and f-minus for A=0.9}, we plot the real part of eigenfrequency of $m=2~f$-mode 
for fixed degree of differential rotation $A=0.9$. 
\footnote{
Figure \ref{fig:f-plus and f-minus for A=0.9} is compared to figure 1 (left) of \citet{Passamonti_Andersson2015} except that we assign a minus sign to the frequency of the retrograde mode.
}
The shaded region is the corotation band of frequency. 
The curve `$f_+$' corresponds to the $f$-mode we consider in this paper. The azimuthal pattern speed of it is in the same direction as the rotation of the background star. Thus the mode pattern is prograde with respect to the stellar rotation. The frequency enters the corotation band as we increase $T/W$ and the mode becomes dynamically unstable (outside the corotation band, the frequency is purely real). On the other hand the mode $f_-$ has the azimuthal pattern speed
in the opposite direction to the stellar rotation. The mode pattern is retrograde. For the classical bar mode instability to occur, the frequency of retrograde $f$-mode needs to go above zero and to merge with the prograde $f_+$-mode. Apparently it is not the case with this instability.

Our 1D model suggests that a possible mechanism of low $T/W$ dynamical instability is the wave over-reflection
at a corotation radius of an eigenmode. 
As in an accretion disc instability studied by \citet{Drury1985} and  by \citet{Tsang_Lai2008}
a wave incident to its corotation radius may be reflected with an amplitude whose modulus exceeds unity.
If the wave is reflected back at edges of the system and a standing-wave condition is satisfied, the
resulting trapped oscillation mode is dynamically unstable. The difference of our study from the disc cases
is that the wave characteristic of our solution is less clear, since the wave number of the unstable modes
here is not high enough so that JWKB treatment is strictly speaking not applicable. With this caution
in mind, we may dare to propose a wave-based depiction of the instability as follows. We see that $K^2$, which may be regarded as a local wave number squared is positive mostly around the surface (Figure \ref{fig:K2 wavenumber}). Thus the wave is trapped mainly in the domain around the surface. The wave travelling inward
is amplified at its corotation radius and reflected back to the surface. The wave is perfectly reflected back again there. If the wave happens to satisfy a standing wave condition, an unstable mode is established with its amplification cycles. This may explain a qualitative agreements of imaginary part of eigenfrequencies  with  
the imaginary part computed by the reflection coefficient at the corotation (Figure \ref{fig:sigi by |R|}).
%

\section{Conclusions}
\label{sec: Conclusions}
In this paper we investigate characteristics of so-called low $T/W$ dynamical instability, especially focusing on the natures of corotation radius of oscillations. When the pattern speed of an oscillation in the azimuthal direction coincides with the rotational angular frequency of a radius inside a star, the equation of linear perturbation formally has a singularity there. Due to the singularity, three type of solution are possible. An oscillation mode may have a complex frequency (growing or damping in time). The eigenfrequency of the mode is discrete which is labelled by an integer number. Or we may have a mode with a purely real frequency which spans a domain of real number. The mode belongs to a continuous spectrum of frequency and is similar to van Kampen modes of plasma oscillations.
These three kinds of modes with corotation radii may coexist in a star with sufficiently high degree of differential rotation.

To see a characteristic of unstable mode with a corotation radius, we study the canonical angular momentum density and its flux. They are defined as quadratic functionals of linear perturbations. Since the canonical angular momentum density satisfies a conservation law (equation (\ref{eq:jc conservation})), its modulus grows in time when the radial derivative of the flux has an opposite sign to it. The distributions of canonical angular momentum density and its flux for unstable mode with corotation radius satisfy this condition around the corotation radius.
We also study the energy flux of wave defined by \citet{Drury1985} for corotating unstable modes. The flux changes its sign at the corotation radius in such a way that the wave is done a work at the corotation radius
to increase its energy.

An over-reflection of wave at a corotation radius is shown to cause dynamical instability for some disc models. We study if a similar mechanism may be at the root of the low $T/W$ dynamical instability of differentially rotating stars. Here we compute the imaginary part of the unstable eigenmodes by assuming the wave trapped between the corotation
point and the surface is over-reflected at the corotation radius. The results qualitatively explains a dependence
of the imaginary part on the parameter to characterize the degree of differential rotation. We argue that corotating unstable $f$-modes with $m\ge 2$ and $p$-modes are destabilized by this mechanism. 

Finally we should also point out that an off-center density maximum (quasi-toroidal density distribution) of a equilibrium star seems not to be related to the appearance of unstable modes. We see unstable modes even if the equilibrium density profile
has a maximum at the origin for stars with weak differential rotation.

\section*{Acknowledgements}
MS is supported in part by JSPS Grant-in-Aid for Young Scientists B (No. 201103201) and the Waseda University Grant for Special Research Projects (2013A-6164).

\bibliographystyle{mnras}

\bibliography{pap}

\begin{thebibliography}{}
\makeatletter
\relax
\def\mn@urlcharsother{\let\do\@makeother \do\$\do\&\do\#\do\^\do\_\do\%\do\~}
\def\mn@doi{\begingroup\mn@urlcharsother \@ifnextchar [ {\mn@doi@}
  {\mn@doi@[]}}
\def\mn@doi@[#1]#2{\def\@tempa{#1}\ifx\@tempa\@empty \href
  {http://dx.doi.org/#2} {doi:#2}\else \href {http://dx.doi.org/#2} {#1}\fi
  \endgroup}
\def\mn@eprint#1#2{\mn@eprint@#1:#2::\@nil}
\def\mn@eprint@arXiv#1{\href {http://arxiv.org/abs/#1} {{\tt arXiv:#1}}}
\def\mn@eprint@dblp#1{\href {http://dblp.uni-trier.de/rec/bibtex/#1.xml}
  {dblp:#1}}
\def\mn@eprint@#1:#2:#3:#4\@nil{\def\@tempa {#1}\def\@tempb {#2}\def\@tempc
  {#3}\ifx \@tempc \@empty \let \@tempc \@tempb \let \@tempb \@tempa \fi \ifx
  \@tempb \@empty \def\@tempb {arXiv}\fi \@ifundefined
  {mn@eprint@\@tempb}{\@tempb:\@tempc}{\expandafter \expandafter \csname
  mn@eprint@\@tempb\endcsname \expandafter{\@tempc}}}

\bibitem[\protect\citeauthoryear{{Baiotti}, {de Pietri}, {Manca}  \&
  {Rezzolla}}{{Baiotti} et~al.}{2007}]{Baiotti_etal2007}
{Baiotti} L.,  {de Pietri} R.,  {Manca} G.~M.,   {Rezzolla} L.,  2007, \mn@doi
  [\prd] {10.1103/PhysRevD.75.044023}, \href
  {http://ads.nao.ac.jp/abs/2007PhRvD..75d4023B} {75, 044023}

\bibitem[\protect\citeauthoryear{{Balbinski}}{{Balbinski}}{1984}]{Balbinski1984a}
{Balbinski} E.,  1984, \mn@doi [\mnras] {10.1093/mnras/209.2.145}, \href
  {http://ads.nao.ac.jp/abs/1984MNRAS.209..145B} {209, 145}

\bibitem[\protect\citeauthoryear{{Balbinski}}{{Balbinski}}{1985}]{Balbinski1985}
{Balbinski} E.,  1985, \mn@doi [\mnras] {10.1093/mnras/216.4.897}, \href
  {http://ads.nao.ac.jp/abs/1985MNRAS.216..897B} {216, 897}

\bibitem[\protect\citeauthoryear{{Blaes} \& {Glatzel}}{{Blaes} \&
  {Glatzel}}{1986}]{Blaes_Glatzel1986}
{Blaes} O.~M.,  {Glatzel} W.,  1986, \mn@doi [\mnras]
  {10.1093/mnras/220.2.253}, \href
  {http://adsabs.harvard.edu/abs/1986MNRAS.220..253B} {220, 253}

\bibitem[\protect\citeauthoryear{{Cairns}}{{Cairns}}{1979}]{Cairns1979}
{Cairns} R.~A.,  1979, \mn@doi [Journal of Fluid Mechanics]
  {10.1017/S0022112079000495}, \href
  {http://ads.nao.ac.jp/abs/1979JFM....92....1C} {92, 1}

\bibitem[\protect\citeauthoryear{{Centrella}, {New}, {Lowe}  \&
  {Brown}}{{Centrella} et~al.}{2001}]{Centrella_etal2001}
{Centrella} J.~M.,  {New} K.~C.~B.,  {Lowe} L.~L.,   {Brown} J.~D.,  2001,
  \mn@doi [\apjl] {10.1086/319634}, \href
  {http://ads.nao.ac.jp/abs/2001ApJ...550L.193C} {550, L193}

\bibitem[\protect\citeauthoryear{{Corvino}, {Rezzolla}, {Bernuzzi}, {De Pietri}
   \& {Giacomazzo}}{{Corvino} et~al.}{2010}]{Corvino_etal2010}
{Corvino} G.,  {Rezzolla} L.,  {Bernuzzi} S.,  {De Pietri} R.,   {Giacomazzo}
  B.,  2010, \mn@doi [Classical and Quantum Gravity]
  {10.1088/0264-9381/27/11/114104}, \href
  {http://ads.nao.ac.jp/abs/2010CQGra..27k4104C} {27, 114104}

\bibitem[\protect\citeauthoryear{{Dirac}}{{Dirac}}{1947}]{Dirac1947}
{Dirac} P.~A.~M.,  1947, {The principles of quantum mechanics, Sec.15}.
Clarendon Press, Oxford

\bibitem[\protect\citeauthoryear{{Drury}}{{Drury}}{1985}]{Drury1985}
{Drury} L.~O.~C.,  1985, \mn@doi [\mnras] {10.1093/mnras/217.4.821}, \href
  {http://ads.nao.ac.jp/abs/1985MNRAS.217..821D} {217, 821}

\bibitem[\protect\citeauthoryear{{Friedman} \& {Schutz}}{{Friedman} \&
  {Schutz}}{1978}]{Friedman_Schutz1978a}
{Friedman} J.~L.,  {Schutz} B.~F.,  1978, \mn@doi [\apj] {10.1086/156098},
  \href {http://ads.nao.ac.jp/abs/1978ApJ...221..937F} {221, 937}

\bibitem[\protect\citeauthoryear{{Goldreich} \& {Narayan}}{{Goldreich} \&
  {Narayan}}{1985}]{Goldreich_Narayan1985}
{Goldreich} P.,  {Narayan} R.,  1985, \mn@doi [\mnras]
  {10.1093/mnras/213.1.7P}, \href
  {http://adsabs.harvard.edu/abs/1985MNRAS.213P...7G} {213, 7}

\bibitem[\protect\citeauthoryear{{Goldreich}, {Goodman}  \&
  {Narayan}}{{Goldreich} et~al.}{1986}]{Goldreich_etal1986_papI}
{Goldreich} P.,  {Goodman} J.,   {Narayan} R.,  1986, \mn@doi [\mnras]
  {10.1093/mnras/221.2.339}, \href
  {http://adsabs.harvard.edu/abs/1986MNRAS.221..339G} {221, 339}

\bibitem[\protect\citeauthoryear{{Goodman}, {Narayan}  \&
  {Goldreich}}{{Goodman} et~al.}{1987}]{Goodman_etal1987_papII}
{Goodman} J.,  {Narayan} R.,   {Goldreich} P.,  1987, \mn@doi [\mnras]
  {10.1093/mnras/225.3.695}, \href
  {http://adsabs.harvard.edu/abs/1987MNRAS.225..695G} {225, 695}

\bibitem[\protect\citeauthoryear{{Hachisu}}{{Hachisu}}{1986}]{HSCF}
{Hachisu} I.,  1986, \mn@doi [\apjs] {10.1086/191121}, \href
  {http://cdsads.u-strasbg.fr/abs/1986ApJS...61..479H} {61, 479}

\bibitem[\protect\citeauthoryear{{Hindmarsh}}{{Hindmarsh}}{1983}]{odepack}
{Hindmarsh} A.~C.,  1983, {in Stepleman R.~S. et al., eds., Scientific
  Computing (Vol.1 of IMACS Transactions on Scientific Computation)}.
North-Holland, Amsterdam, p.55

\bibitem[\protect\citeauthoryear{{Ipser} \& {Lindblom}}{{Ipser} \&
  {Lindblom}}{1990}]{IpserLindblom2pot1990}
{Ipser} J.~R.,  {Lindblom} L.,  1990, \mn@doi [\apj] {10.1086/168757}, \href
  {http://ads.nao.ac.jp/abs/1990ApJ...355..226I} {355, 226}

\bibitem[\protect\citeauthoryear{{Karino} \& {Eriguchi}}{{Karino} \&
  {Eriguchi}}{2003}]{Karino_Eriguchi2003}
{Karino} S.,  {Eriguchi} Y.,  2003, \mn@doi [\apj] {10.1086/375768}, \href
  {http://ads.nao.ac.jp/abs/2003ApJ...592.1119K} {592, 1119}

\bibitem[\protect\citeauthoryear{{Kato}}{{Kato}}{2001}]{Kato2001}
{Kato} S.,  2001, \mn@doi [\pasj] {10.1093/pasj/53.1.1}, \href
  {http://adsabs.harvard.edu/abs/2001PASJ...53....1K} {53, 1}

\bibitem[\protect\citeauthoryear{Kuroda, Takiwaki  \& Kotake}{Kuroda
  et~al.}{2014}]{Kuroda_etal2014}
Kuroda T.,  Takiwaki T.,   Kotake K.,  2014, \mn@doi [Phys. Rev. D]
  {10.1103/PhysRevD.89.044011}, 89, 044011

\bibitem[\protect\citeauthoryear{{Lehner}, {Liebling}, {Palenzuela}  \&
  {Motl}}{{Lehner} et~al.}{2016}]{Lehner_etal2016}
{Lehner} L.,  {Liebling} S.~L.,  {Palenzuela} C.,   {Motl} P.~M.,  2016,
  \mn@doi [\prd] {10.1103/PhysRevD.94.043003}, \href
  {http://adsabs.harvard.edu/abs/2016PhRvD..94d3003L} {94, 043003}

\bibitem[\protect\citeauthoryear{{Lockitch}, {Andersson}  \&
  {Watts}}{{Lockitch} et~al.}{2004}]{Lockitch_etal2004}
{Lockitch} K.~H.,  {Andersson} N.,   {Watts} A.~L.,  2004, \mn@doi [Classical
  and Quantum Gravity] {10.1088/0264-9381/21/19/012}, \href
  {http://ads.nao.ac.jp/abs/2004CQGra..21.4661L} {21, 4661}

\bibitem[\protect\citeauthoryear{{Luyten}}{{Luyten}}{1990}]{Luyten1990}
{Luyten} P.~J.,  1990, \mnras, \href
  {http://ads.nao.ac.jp/abs/1990MNRAS.245..614L} {245, 614}

\bibitem[\protect\citeauthoryear{{Narayan}, {Goldreich}  \&
  {Goodman}}{{Narayan} et~al.}{1987}]{Narayan_etal1987}
{Narayan} R.,  {Goldreich} P.,   {Goodman} J.,  1987, \mn@doi [\mnras]
  {10.1093/mnras/228.1.1}, \href
  {http://adsabs.harvard.edu/abs/1987MNRAS.228....1N} {228, 1}

\bibitem[\protect\citeauthoryear{Olver, Lozier, Boisvert  \& Clark}{Olver
  et~al.}{2010}]{NISTHandbook}
Olver F.~W.,  Lozier D.~W.,  Boisvert R.~F.,   Clark C.~W.,  2010, NIST
  Handbook of Mathematical Functions, Sec.2.7.
Cambridge University Press, New York, NY, USA

\bibitem[\protect\citeauthoryear{{Ostriker} \& {Bodenheimer}}{{Ostriker} \&
  {Bodenheimer}}{1973}]{Ostriker_Bodenheimer1973}
{Ostriker} J.~P.,  {Bodenheimer} P.,  1973, \mn@doi [\apj] {10.1086/151952},
  \href {http://ads.nao.ac.jp/abs/1973ApJ...180..171O} {180, 171}

\bibitem[\protect\citeauthoryear{{Ostriker} \& {Tassoul}}{{Ostriker} \&
  {Tassoul}}{1969}]{Ostriker_Tassoul1969}
{Ostriker} J.~P.,  {Tassoul} J.~L.,  1969, \mn@doi [\apj] {10.1086/149927},
  \href {http://ads.nao.ac.jp/abs/1969ApJ...155..987O} {155, 987}

\bibitem[\protect\citeauthoryear{{Ott}, {Ou}, {Tohline}  \& {Burrows}}{{Ott}
  et~al.}{2005}]{Ott_etal2005}
{Ott} C.~D.,  {Ou} S.,  {Tohline} J.~E.,   {Burrows} A.,  2005, \mn@doi [\apjl]
  {10.1086/431305}, \href {http://ads.nao.ac.jp/abs/2005ApJ...625L.119O} {625,
  L119}

\bibitem[\protect\citeauthoryear{{Ou} \& {Tohline}}{{Ou} \&
  {Tohline}}{2006}]{Ou_Tohline2006}
{Ou} S.,  {Tohline} J.~E.,  2006, \mn@doi [\apj] {10.1086/507597}, \href
  {http://ads.nao.ac.jp/abs/2006ApJ...651.1068O} {651, 1068}

\bibitem[\protect\citeauthoryear{{Papaloizou} \& {Pringle}}{{Papaloizou} \&
  {Pringle}}{1984}]{Papaloizou_Pringle1984}
{Papaloizou} J.~C.~B.,  {Pringle} J.~E.,  1984, \mn@doi [\mnras]
  {10.1093/mnras/208.4.721}, \href
  {http://ads.nao.ac.jp/abs/1984MNRAS.208..721P} {208, 721}

\bibitem[\protect\citeauthoryear{{Passamonti} \& {Andersson}}{{Passamonti} \&
  {Andersson}}{2015}]{Passamonti_Andersson2015}
{Passamonti} A.,  {Andersson} N.,  2015, \mn@doi [\mnras]
  {10.1093/mnras/stu2062}, \href {http://ads.nao.ac.jp/abs/2015MNRAS.446..555P}
  {446, 555}

\bibitem[\protect\citeauthoryear{{Pickett}, {Durisen}  \& {Davis}}{{Pickett}
  et~al.}{1996}]{Picket_etal1996}
{Pickett} B.~K.,  {Durisen} R.~H.,   {Davis} G.~A.,  1996, \mn@doi [\apj]
  {10.1086/176852}, \href {http://ads.nao.ac.jp/abs/1996ApJ...458..714P} {458,
  714}

\bibitem[\protect\citeauthoryear{{Radice}, {Bernuzzi}  \& {Ott}}{{Radice}
  et~al.}{2016}]{Radice_etal2016}
{Radice} D.,  {Bernuzzi} S.,   {Ott} C.~D.,  2016, \mn@doi [\prd]
  {10.1103/PhysRevD.94.064011}, \href
  {http://ads.nao.ac.jp/abs/2016PhRvD..94f4011R} {94, 064011}

\bibitem[\protect\citeauthoryear{{Saijo} \& {Yoshida}}{{Saijo} \&
  {Yoshida}}{2006}]{Saijo_Yoshida2006}
{Saijo} M.,  {Yoshida} S.,  2006, \mn@doi [\mnras]
  {10.1111/j.1365-2966.2006.10229.x}, \href
  {http://ads.nao.ac.jp/abs/2006MNRAS.368.1429S} {368, 1429}

\bibitem[\protect\citeauthoryear{{Saijo} \& {Yoshida}}{{Saijo} \&
  {Yoshida}}{2016}]{Saijo_Yoshida2016}
{Saijo} M.,  {Yoshida} S.,  2016, \mn@doi [\prd] {10.1103/PhysRevD.94.084032},
  \href {http://ads.nao.ac.jp/abs/2016PhRvD..94h4032S} {94, 084032}

\bibitem[\protect\citeauthoryear{{Saijo}, {Shibata}, {Baumgarte}  \&
  {Shapiro}}{{Saijo} et~al.}{2001}]{Saijo_etal2001}
{Saijo} M.,  {Shibata} M.,  {Baumgarte} T.~W.,   {Shapiro} S.~L.,  2001,
  \mn@doi [\apj] {10.1086/319016}, \href
  {http://ads.nao.ac.jp/abs/2001ApJ...548..919S} {548, 919}

\bibitem[\protect\citeauthoryear{{Saijo}, {Baumgarte}  \& {Shapiro}}{{Saijo}
  et~al.}{2003}]{Saijo_etal2003}
{Saijo} M.,  {Baumgarte} T.~W.,   {Shapiro} S.~L.,  2003, \mn@doi [\apj]
  {10.1086/377334}, \href {http://ads.nao.ac.jp/abs/2003ApJ...595..352S} {595,
  352}

\bibitem[\protect\citeauthoryear{{Schutz}}{{Schutz}}{1980}]{Schutz1980_pap3}
{Schutz} B.~F.,  1980, \mn@doi [\mnras] {10.1093/mnras/190.1.21}, \href
  {http://ads.nao.ac.jp/abs/1980MNRAS.190...21S} {190, 21}

\bibitem[\protect\citeauthoryear{{Shibata}, {Baumgarte}  \&
  {Shapiro}}{{Shibata} et~al.}{2000}]{Shibata_etal2000barmode}
{Shibata} M.,  {Baumgarte} T.~W.,   {Shapiro} S.~L.,  2000, \mn@doi [\apj]
  {10.1086/309525}, \href {http://ads.nao.ac.jp/abs/2000ApJ...542..453S} {542,
  453}

\bibitem[\protect\citeauthoryear{{Shibata}, {Karino}  \& {Eriguchi}}{{Shibata}
  et~al.}{2002}]{Shibata_etal2002}
{Shibata} M.,  {Karino} S.,   {Eriguchi} Y.,  2002, \mn@doi [\mnras]
  {10.1046/j.1365-8711.2002.05724.x}, \href
  {http://ads.nao.ac.jp/abs/2002MNRAS.334L..27S} {334, L27}

\bibitem[\protect\citeauthoryear{{Shibata}, {Karino}  \& {Eriguchi}}{{Shibata}
  et~al.}{2003}]{Shibata_etal2003}
{Shibata} M.,  {Karino} S.,   {Eriguchi} Y.,  2003, \mn@doi [\mnras]
  {10.1046/j.1365-8711.2003.06699.x}, \href
  {http://ads.nao.ac.jp/abs/2003MNRAS.343..619S} {343, 619}

\bibitem[\protect\citeauthoryear{{Stix}}{{Stix}}{1992}]{Stix_waves_in_plasmas}
{Stix} T.~H.,  1992, { Waves in plasmas, Chap.8}.
American Institute of Physics, New York

\bibitem[\protect\citeauthoryear{{Takiwaki}, {Kotake}  \& {Suwa}}{{Takiwaki}
  et~al.}{2016}]{Takiwaki_etal2016}
{Takiwaki} T.,  {Kotake} K.,   {Suwa} Y.,  2016, \mn@doi [\mnras]
  {10.1093/mnrasl/slw105}, \href {http://ads.nao.ac.jp/abs/2016MNRAS.461L.112T}
  {461, L112}

\bibitem[\protect\citeauthoryear{{Tassoul}}{{Tassoul}}{1978}]{Tassoul1978book}
{Tassoul} J.-L.,  1978, {Theory of rotating stars, Chap.10}.
Princeton University Press, New Jersey

\bibitem[\protect\citeauthoryear{{Tassoul} \& {Ostriker}}{{Tassoul} \&
  {Ostriker}}{1968}]{Tassoul_Ostriker1968}
{Tassoul} J.~L.,  {Ostriker} J.~P.,  1968, \mn@doi [\apj] {10.1086/149784},
  \href {http://ads.nao.ac.jp/abs/1968ApJ...154..613T} {154, 613}

\bibitem[\protect\citeauthoryear{{Toman}, {Imamura}, {Pickett}  \&
  {Durisen}}{{Toman} et~al.}{1998}]{Toman_etal1998}
{Toman} J.,  {Imamura} J.~N.,  {Pickett} B.~K.,   {Durisen} R.~H.,  1998,
  \mn@doi [\apj] {10.1086/305466}, \href
  {http://ads.nao.ac.jp/abs/1998ApJ...497..370T} {497, 370}

\bibitem[\protect\citeauthoryear{{Tsang} \& {Lai}}{{Tsang} \&
  {Lai}}{2008}]{Tsang_Lai2008}
{Tsang} D.,  {Lai} D.,  2008, \mn@doi [\mnras]
  {10.1111/j.1365-2966.2008.13252.x}, \href
  {http://ads.nao.ac.jp/abs/2008MNRAS.387..446T} {387, 446}

\bibitem[\protect\citeauthoryear{{Vallis}}{{Vallis}}{2006}]{Vallis2006book}
{Vallis} G.~K.,  2006, {Atmospheric and Oceanic Fluid Dynamics, Chap.6}.
Cambridge University Press, Cambridge

\bibitem[\protect\citeauthoryear{{Watts}, {Andersson}, {Beyer}  \&
  {Schutz}}{{Watts} et~al.}{2003}]{Watts_etal2003}
{Watts} A.~L.,  {Andersson} N.,  {Beyer} H.,   {Schutz} B.~F.,  2003, \mn@doi
  [\mnras] {10.1046/j.1365-8711.2003.06612.x}, \href
  {http://ads.nao.ac.jp/abs/2003MNRAS.342.1156W} {342, 1156}

\bibitem[\protect\citeauthoryear{{Watts}, {Andersson}  \& {Jones}}{{Watts}
  et~al.}{2005}]{Watts_etal2005}
{Watts} A.~L.,  {Andersson} N.,   {Jones} D.~I.,  2005, \mn@doi [\apjl]
  {10.1086/427653}, \href {http://ads.nao.ac.jp/abs/2005ApJ...618L..37W} {618,
  L37}

\bibitem[\protect\citeauthoryear{{Yoshida}}{{Yoshida}}{2006}]{Yoshida2006}
{Yoshida} S.,  2006, \mn@doi [Classical and Quantum Gravity]
  {10.1088/0264-9381/23/23/018}, \href
  {http://ads.nao.ac.jp/abs/2006CQGra..23.6899Y} {23, 6899}

\bibitem[\protect\citeauthoryear{{Yoshida} \& {Eriguchi}}{{Yoshida} \&
  {Eriguchi}}{1995}]{Yoshida_Eriguchi1995}
{Yoshida} S.,  {Eriguchi} Y.,  1995, \mn@doi [\apj] {10.1086/175126}, \href
  {http://ads.nao.ac.jp/abs/1995ApJ...438..830Y} {438, 830}

\bibitem[\protect\citeauthoryear{{van Kampen}}{{van
  Kampen}}{1955}]{VanKampen1955}
{van Kampen} N.~G.,  1955, \mn@doi [Physica] {10.1016/S0031-8914(55)93068-8},
  \href {http://ads.nao.ac.jp/abs/1955Phy....21..949V} {21, 949}

\makeatother
\end{thebibliography}
%

\appendix
\section{Equations for stellar perturbation in two potential formalism}
\label{appendix:2potentials}
\citet{IpserLindblom2pot1990} wrote the perturbed equations of self-gravitating Newtonian fluid
configuration with stationarity and axi-symmetry, 
in a coupled partial differential equations for two scalar potentials $\delta U$ and $\delta\Phi$ (the variable with $\delta$ means taking its Eulerian perturbation.). We adopt the formalism in the cylindrical coordinate. 
Taking Eulerian perturbation of equation (\ref{eq:Euler0}), we have
	\begin{eqnarray}
		&&\gamma_{ab}\del{t}\delta v^b + (\nabla_b v_a- \nabla_a v_b)\delta v^b 
		+ v^\varphi\del{\varphi}(\gamma_{ab}\delta v^b) \nonumber \\
		&& -v^\varphi\del{a}(\gamma_{\varphi b}\delta v^b) 
		 + \del{a}(\gamma_{\varphi b}v^\varphi \delta v^b) = -\nabla_a\delta U.
	\end{eqnarray}
Here $v^a$ is equilibrium velocity component, i.e., $v^a = \Omega\varphi^a$ and $\varphi^a$ is the coordinate basis vector of $\varphi$, $\varphi^a = \left(\frac{\partial}{\partial\varphi}\right)^a$.
$\gamma_{ab} $ is the spatial metric of the cylindrical coordinate. By assuming harmonic dependence of perturbed variables on $t$ and $\varphi$ as $\delta U\sim \eNap^{-i\sigma t+im\varphi}$, we rewrite the equation as
	\begin{equation}
		is\gamma_{ab}\delta v^b + (\nabla_a v_b- \nabla_b v_a)\delta v^b 
		- \del{a}~\gamma_{\varphi\varphi}\delta v^\varphi = \nabla_a\delta U,
	\end{equation}
	where $s=\sigma-m\Omega$ is  the frequency seen in a local co-moving observer to the equilibrium flow.
The equation is solved for $\delta v^a$ as
	\begin{eqnarray}
		\label{eq:perturbed vR}
		i\delta v^R &=& \frac{s}{L}\del{R}\delta U - \frac{2m\Omega}{RL}\delta U,\\
		\label{eq:perturbed vPhi}
		\delta v^\varphi &=& -\frac{\kappa^2}{2\Omega LR}\del{R}\delta U + \frac{ms}{R^2L}\delta U,\\
		\label{eq:perturbed vZ}
		i\delta v^z &=& \frac{1}{s}\del{z}\delta U,	
			\end{eqnarray}
where $\kappa^2= 2\Omega\left(2\Omega + R\del{R}\Omega\right)$ is epicyclic frequency squared
and $L=s^2-\kappa^2$. Perturbed continuity equation is obtained from equation (\ref{eq:continuity0}),
	\begin{equation}
		s\delta\rho+ \frac{1}{R}\del{R}\left(R\rho(i\delta v^R)\right) - m\rho\delta v^\varphi + \del{z}(\rho\delta v^z) = 0.
		\end{equation}
Substituting equations (\ref{eq:perturbed vR}), (\ref{eq:perturbed vPhi}) and (\ref{eq:perturbed vZ}) to this equation, we obtain a second order differential equation for one of the scalar
potentials $\delta U$ as
	\begin{eqnarray}
	\label{eq:PDE_dU}
		&&\frac{s}{R}\del{R}\left[R\rho\left(\frac{s}{L}\del{R}\delta U-\frac{2m\Omega}{RL}\delta U\right)\right] \nonumber \\
		&&-m\rho\left(-\frac{s\kappa^2}{2\Omega LR}\del{R}\delta U+\frac{ms^2}{R^2L}\delta U\right) \nonumber \\
		&&+ s\del{z}\left(\frac{\rho}{s}\del{z}\delta U\right)
		+ s^2 \rho\frac{d\rho}{dp}(\delta U-\delta\Phi) = 0.
		\end{eqnarray}
Here $\delta\rho = \rho\frac{d\rho}{dp}(\delta U-\delta\Phi)$ is used.

Poisson equation for perturbed gravitational potential $\delta\Phi$ is
	\begin{equation}
		\label{eq:PDE_dPhi}	
		\left(\del{R}^2+\frac{1}{R}\del{R}-\frac{m^2}{R^2}\right)\delta\Phi = 4\pi\rho\frac{d\rho}{dp}(\delta U-\delta\Phi).
		\end{equation}

At a Lindblad radius ($L(R)=s^2-\kappa(R)^2=0$),  equation (\ref{eq:PDE_dU}) has an apparently singular coefficients. 
These terms are, however, regular there. They are combined as
	\begin{equation}
		\frac{\del{R}L}{L}\left(\del{R}L-\frac{2m\Omega}{sR}\delta U\right).
	\end{equation}
By equations (\ref{eq:perturbed vR}) and (\ref{eq:perturbed vPhi}) we see $\del{R}\delta U-\frac{2m\Omega}{sR}\delta U$ needs to vanish as $L\to 0$ in order for the velocity perturbations to be regular there.

\section{Frobenius series solution around the corotation radius}
\label{appendix:corotation}
With a Laurent series expansion of the coefficients,  the equation for $\delta U$ (equation (\ref{eq:2potentials_dU})) may be written near a corotation radius
of oscillation $\sigma-m\Omega(\Rc)=0$ as
	\begin{equation}
		\frac{d^2}{dx^2}\delta U + p(x)\frac{d}{dx}\delta U + q(x)\delta U=0,
	\label{eq:deltaUeq_appendix}
	\end{equation}
		with $x=R-\Rc$ where 
	\begin{eqnarray}
		&&p(x)=p_{_0} + p_{_1} x  + {\cal O}[x^2],\\
		&&q(x)=q_{_{-1}}x^{-1}+q_{_0} + {\cal O}[x].
	\end{eqnarray}
Here $p_{_i}$, $q_{_i}$ are constants computed from the equilibrium quantities at $R=\Rc$,
	\begin{eqnarray}
		p_{_0} &=& \del{R}\ln\left(\frac{R\rho}{L}\right),\\
		p_{_1} &=& -\frac{1}{R^2} -\frac{\rho\del{R}^2\rho-(\del{R}\rho)^2}{\rho^2}
		-\frac{L\del{R}^2L-(\del{R}L)^2}{L^2},\\
		q_{_{-1}} &=& \frac{2}{-R\del{R}(\ln\Omega)}\del{R}\ln\left(\frac{L}{\rho\Omega}\right),\\
		q_{_0} &=& \frac{L}{c_{_\mathrm{eq}}^2} - \frac{m^2}{R^2}
		+ \frac{2}{-R\del{R}(\ln\Omega)}
			\left[
				\left(-\frac{\del{R}^2\Omega}{2\Omega} -\frac{1}{R}\right)
				  \del{R}\ln\left(\frac{L}{\rho\Omega}\right)  \right. \nonumber\\
		      &&    \left. +\frac{L\del{R}^2L-(\del{R}L)^2}{L^2}
		      			- \frac{\rho\del{R}^2\rho-(\del{R}\rho)^2}{\rho^2}  \right.\nonumber\\
					&&  \left.  - \frac{\Omega\del{R}^2\Omega-(\del{R}\Omega)^2}{\Omega^2} 
			\right].
	\end{eqnarray}

We here omit the perturbation of gravitational potential since it is regular and small compared to other terms
of the equation. A formal series solution of a form $\delta U = \Sigma_{n=0}^\infty a_nx^{\lambda+n}$ is obtained
by substituting the series to equation (\ref{eq:deltaUeq_appendix}) and compare each terms with a same power of $x$  \citep{NISTHandbook}.
The coefficients of the series solve a recursive relation ($k=1,2,\cdots$),
	\begin{equation}
		Q(\lambda+k)a_k = - \sum_{j=0}^{k-1}[(\lambda + j) p_{_{k-1-j}}+q_{_{k-2-j}}]a_j,
	\end{equation}
where $Q(\lambda)$ defines the indicial equation that reads $Q(\lambda)=\lambda(\lambda-1)=0$ in our case. Thus one of the series solutions for $\lambda=1$, $y_1(x)$, is completely regular with the lowest order term proportional to  $x^1$. 
Another linear independent solution is obtained by formally solving the recursive equation as a function
of the parameter $\lambda$ and taking the limit as
	\begin{equation}
y_2(x) = \lim_{\lambda\to 0}\frac{\partial}{\partial \lambda}
\left[
  \lambda(a_0(\lambda)x^\lambda+a_1(\lambda)x^{\lambda+1} + O(x^{\lambda + 2}))
\right].
	\end{equation}
$y_2(x)$ is a summation of a regular series term starting with the order $x^0$ and a term as $y_1(x)\ln x$. 
The linearly independent basis $(y_1(x), y_2(x))$ for equation (\ref{eq:deltaUeq_appendix}) is thus fixed by giving
$p_i$ and $q_i$ at $R=\Rc$. Then the general solution equation (\ref{eq:deltaUeq_appendix}) is written as
	\begin{equation}
		\delta U(x) = K_1 y_1(x) + K_2 y_2(x),
	\end{equation}
where $K_1, K_2$ are complex constants. In the numerical bridging of section \ref{sec:bridging_explained}, these constants are determined near the corotation point $R=\Rc - \epsC$, by matching it with the numerical
solution there.
\section{Canonical angular momentum and its flux}
\label{sec:jc and flux}
\citet{Friedman_Schutz1978a} developed a Lagrangian perturbation theory of non-relativistic fluid. They
introduced conserved canonical quantities, which we utilize for our normal mode analysis.

Lagrangian displacement vector $\xi^i$ ($i$ is a spatial coordinate index) is defined as a vector field
connecting identical fluid elements of equilibrium and perturbed state. By using perturbed Euler equation,
Poisson's equation and equation of continuity, the perturbed fluid is described by an differential equation
of $\xi^i$,
	\begin{equation}
		A^i_j\del{t}^2\xi^j + B^i_j\del{t}\xi^j + C^i_j\xi^j = 0,
		\label{eq:xi-equation}
	\end{equation}
where $A^i_j, B^i_j,C^i_j$ are differential operators. For the detailed expressions of these operators, see equations (32) to (34) of \citet{Friedman_Schutz1978a}.  An inner product of two displacement vectors $\eta$ and $\xi$ is defined as
	\begin{equation}
		\langle\eta_i,\xi^i\rangle \equiv \int_\Sigma \eta^*_i\xi^i dV,
		\end{equation}
where $\Sigma$ is the spatial domain occupied by the fluid (here the asterisk `*' means taking complex conjugate). 
The operators are Hermitian, anti-Hermitian and Hermitian respectively:
	\begin{eqnarray}
		&&\langle\eta_i,A_j\xi^j\rangle = \langle\xi_i,A^i_i\eta^j\rangle^*, \\
		&&\langle\eta_i,B_j\xi^j\rangle = -\langle\xi_i,B^i_i\eta^i\rangle^*,\\
		&&\langle\eta_i,C_j\xi^j\rangle = \langle\xi_i,C^i_i\eta^i\rangle^*.
			\end{eqnarray}

The dynamical equation for displacement equation (\ref{eq:xi-equation}) is derived from a variational principle whose
Lagrangian density is,
	\begin{equation}
		{\cal L} = \frac{1}{2}\left[\del{t}\xi_i A^i_j\del{t}\xi^j + \del{t}\xi_i B^i_j\xi^j - \xi_i C^i_j\xi^i\right].
		\end{equation}

By using the associated symplectic structure to the Lagrangian, conserved canonical energy and angular momentum
are defined. Canonical angular momentum $J_\mathrm{c}$ is written as
	\begin{equation}
		J_\mathrm{c} = \Re\left[\langle\del{\varphi}\xi_i, ~A^i_j\del{t}\xi^j+\frac{1}{2}B^i_j\xi^j\rangle\right].
		\label{eq:canonical angular momentum}
		\end{equation}
For an axisymmetric equilibrium, $J_\mathrm{c}$ is conserved, thus $\frac{d}{dt}J_\mathrm{c}=0$. 
In this case we may assume the perturbed variables to have a harmonic dependence as $\xi \propto \eNap^{im\varphi}$.
Differentiating equation (\ref{eq:canonical angular momentum}) with respect to time, we obtain
	\begin{equation}
		\frac{d}{dt}J_\mathrm{c} = m \Im[\langle\xi_i,C^i_j\xi^j\rangle],
		\label{eq:dJdt}
		\end{equation}
where we use the anti-Hermitian nature of operator $B^i_j$ and the dynamical equation (equation (\ref{eq:xi-equation})).
The quadratic form on the right hand side of equation (\ref{eq:dJdt}) contains such purely real terms as $-\rho (v^k\nabla_k\xi_j^*)(v^k\nabla_k\xi^j)$ which we drop in computing it (here $\rho$ and $v^k$ are the equilibrium density and velocity. $\nabla$ is the spatial covariant derivative). The integrand of equation (\ref{eq:canonical angular momentum}) which we call canonical angular momentum density, $\jc$, is explicitly written as
	\begin{equation}
		j_\mathrm{c} = m\rho \Re[\sigma-m\Omega]|\xi|^2 - mR\Omega \Im[\xi^{\varphi*}\xi^R-\xi^{r*}\xi^\varphi].
		\label{eq:jc defined}
		\end{equation}
The time derivative of $j_\mathrm{c}$ is written as
	\begin{equation}
		\del{t}\jc = \Im[\xi_i^*C^i_j\xi^j] = \nabla_i (-f^i),
		\label{eq:dtjc}
	\end{equation}
where $f^i$ is
	\begin{eqnarray}
		f^i &=& m\Im\left[-\rho v^i\xi_k^*(v\cdot\nabla)\xi^k + \gamma p\xi^{i*}\xi^k\nabla_k p\right. \nonumber \\
			&&\left.- \delta\Phi\left(\frac{1}{4\pi}\delta\Phi^* + \rho\xi^{i*}\right)\right].
		\label{eq:jflux}
		\end{eqnarray}
Here $p$ means the pressure of equilibrium flow. $\gamma$ is the adiabatic index of the gas.

\bsp	

\label{lastpage}

\end{document}